\documentclass[reprint,preprintnumbers,superscriptaddress,longbibliography]{revtex4-2}
\usepackage{orcidlink}
\usepackage{graphicx}
\usepackage{dcolumn}
\usepackage{bm}
\usepackage{textgreek}
\usepackage{placeins}
\usepackage{braket}
\usepackage{physics}
\usepackage{overpic}
\usepackage{amssymb}
\usepackage{amsmath}
\usepackage{subfigure}
\usepackage{qcircuit}
\usepackage{tikz}
\usepackage{adjustbox}
\usepackage{longtable}
\usepackage{soul}

\begin{document}

\title{String Breaking in the Heavy Quark Limit with Scalable Circuits}

\author{Anthony N. Ciavarella \,\orcidlink{0000-0003-3918-4110}}
\email{anciavarella@lbl.gov}
\affiliation{Physics Division, Lawrence Berkeley National Laboratory, Berkeley, California 94720, USA}

\date{\today}

\begin{abstract}
Quantum simulations of non-Abelian gauge theories require efficient mappings onto quantum computers and practical state preparation and measurement procedures. A truncation of the Hilbert space of non-Abelian lattice gauge theories with matter in the heavy quark limit is developed. This truncation is applied to $SU(2)$ lattice gauge theory in $1+1D$ to map the theory efficiently onto a quantum computer. Scalable variational circuits are found to prepare the vacuum and single meson states. It is also shown how these state preparation circuits can be used to perform measurements of the number of mesons produced during the system's time evolution. A state with a single $q\Bar{q}$ pair is prepared on quantum hardware and the inelastic production of $q\Bar{q}$ pairs is observed using $104$ qubits on IBM's Heron quantum computer {\tt ibm\_torino}.

\end{abstract}

\maketitle

\section{Introduction}
High energy physics is directly probed at particle colliders where large numbers of quark anti-quark pairs are produced. As these particles propagate, they bind together into hadrons. Semi-classical models fit to empirical data can be used to describe this process, but observations of collective behavior in proton-proton collisions suggest that the semi-classical limit may be inadequate~\cite{andersson1998lund,bozek2013collective}. This process of hadronization involves the non-perturbative dynamics of quantum chromodynamics (QCD). The non-perturbative regime of QCD can be studied numerically through the use of a lattice~\cite{wilson1974confinement}. Lattice QCD studies on traditional computers have successfully described many static properties of QCD~\cite{,borsanyi2015ab,karsch2003hadron,tiburzi2017double,beane2015ab,savage2017proton}. However, direct studies of the dynamics of lattice QCD suffer from a sign problem on classical computers and require exponentially scaling resources.

The potential of quantum computers to directly probe the dynamics of lattice QCD has inspired a large body of work studying the simulation of gauge theories on quantum computers~\cite{feynman2018simulating,Bauer_2023,humble2022snowmass,humble2022snowmass2,beck2023quantum,dimeglio2023quantum,nielsen2001quantum}. Exploratory simulations of small systems have been performed on existing quantum hardware~\cite{martinez2016real,klco20202,Rahman:2022rlg,rahman2022real,ciavarella2021trailhead,ciavarella2022preparation,alam2022primitive,Illa:2022jqb,Gustafson_2022,Atas_2021,farrell2023preparations,farrell2023preparations2,yang2020observation,Zhou_2022,Su_2023,zhang2023observation,mildenberger2022probing,ciavarella2023quantum,farrell2023scalable,Farrell:2024fit,charles2023simulating,kavaki2024square,mueller2022quantum,ciavarella2024quantum,turro2024classical,de2024observationstring,guo2024concurrent,halimeh2024disorder,li2024sequency,Turro:2024shh,Hayata:2024fnh}. Additionally, a few quantum simulations in $1+1D$ have been performed at scale, demonstrating how the techniques developed for small systems can be used at scale~\cite{farrell2023scalable,Farrell:2024fit,Zemlevskiy:2024vxt}. Most work studying the mapping of gauge theories onto quantum computers has focused on discretizing gauge fields to map them onto qubits. Truncation strategies have been developed for both electric~\cite{byrnes2006simulating,zohar2012simulating,zohar2013quantum,zohar2013cold,zohar2015formulation,zohar2015quantum,zohar2022quantum,raychowdhury2020solving,raychowdhury2020loop,kadam2023loop,shaw2020quantum,ciavarella2022conceptual,ciavarella2022preparation,klco20202,ciavarella2021trailhead,kavaki2024square,rahman2022real,Atas_2021,Rahman:2022rlg,paulson2021simulating,halimeh2023spin,meurice2021theoretical,davoudi2020towards,davoudi2021search,belyansky2023highenergy,berenstein2023integrable,rigobello2023hadrons,kane2024nearlyoptimal,hariprakash2023strategies,su2024coldatom,ciavarella2024quantum} and magnetic bases~\cite{alexandru2019gluon,lamm2019general,ji2020gluon,alexandru2022spectrum,alam2022primitive,Gustafson_2022,zache2023fermionqudit,gonzalez2022hardware}. Encoding a theory onto a quantum computer is only the first step in performing a simulation. Physically relevant states such as the vacuum need to be prepared. Both adiabatic~\cite{jordan2011quantum,jordan7115quantum,preskill2018simulating,jordan2014quantum,jordan2012quantum,chakraborty2020classically,li2024sequency} and variational~\cite{Atas_2021,ciavarella2022preparation,paulson2021simulating,farrell2023scalable} techniques have been explored for preparing vacuum states. Various methods have also been proposed for preparing excited states~\cite{jordan2011quantum,jordan7115quantum,preskill2018simulating,jordan2014quantum,jordan2012quantum,Farrell:2024fit,guo2024concurrent,xu2023concurrent,kane2024nearlyoptimal,Zemlevskiy:2024vxt}. Once a relevant state has been prepared, it needs to be evolved in time and measured.

In this work, a truncation based on the heavy quark limit is developed. In traditional lattice QCD, simulations with heavier than physical quarks have been found to be computationally cheaper. Physics at physical quark masses can be extracted by performing extrapolations to the physical quark mass. Similar benefits are found for performing quantum simulations in the heavy quark limit. It is explicitly demonstrated how this truncation can be used to map a $1+1D$ $SU(2)$ lattice gauge theory onto a quantum computer. Scalable variational circuits are designed to prepare the vacuum and single meson states of this theory on a quantum computer. It is also shown how to use these circuits to perform measurements of the propagation of mesons on the lattice. Using this formalism, a state with a single $q\Bar{q}$ pair is prepared on a quantum computer. The $q\Bar{q}$ pair is evolved in time and the dynamics of the resulting mesons are studied.

\section{Heavy Quark Truncation}
Lattice QCD contains both fermionic quark degrees of freedom and bosonic gauge fields. Quantum computers have a finite Hilbert space which requires the truncation of gauge fields. Note that in one spatial dimension, it is possible to completely integrate out the gauge fields. As fermions are described by a finite-dimensional Hilbert space, it is not strictly necessary to truncate the quark degrees of freedom when encoding the theory onto a quantum computer. However, truncation of the quark degrees of freedom may be beneficial when exploring certain parameter regimes. For example, one can consider working in the heavy quark limit. This regime of QCD is described by Heavy Quark Effective Theory (HQET) and is used to describe the dynamics of mesons containing heavy quarks such as the bottom and charm quarks~\cite{Manohar:2000dt,Korner:1991kf,Isgur:1989vq}. Additionally, lattice QCD on traditional supercomputers has historically been performed with heavier than physical quarks. This led to reductions in computational costs and results at physical quark masses were obtained through extrapolations. Working in this limit when performing quantum simulations of lattice QCD may have similar benefits. 

\begin{figure}
    \centering
    \includegraphics[width=8.6cm]{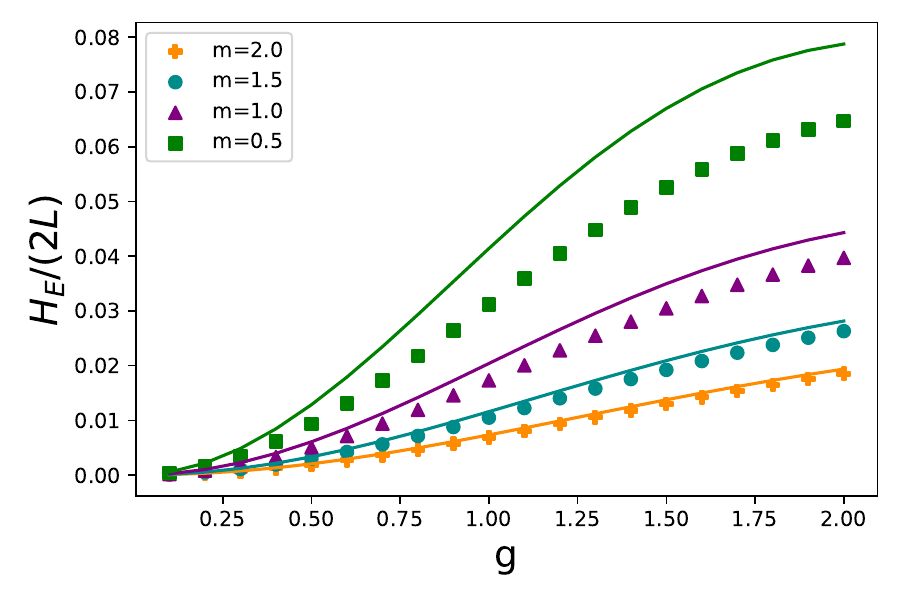}
\caption{Expectation of the electric energy in the vacuum state as a function of $g$ for different quark masses on a lattice with $3$ physical sites. The solid lines were computed using the untruncated Hamiltonian in Eq.~\eqref{eq:QCD1DGauge} and the points were computed by projecting out all states with multiple (anti)quarks on a single site.}
    \label{fig:heavy_electric_exp}
\end{figure}

\begin{figure}
    \centering
    \includegraphics[width=8.6cm]{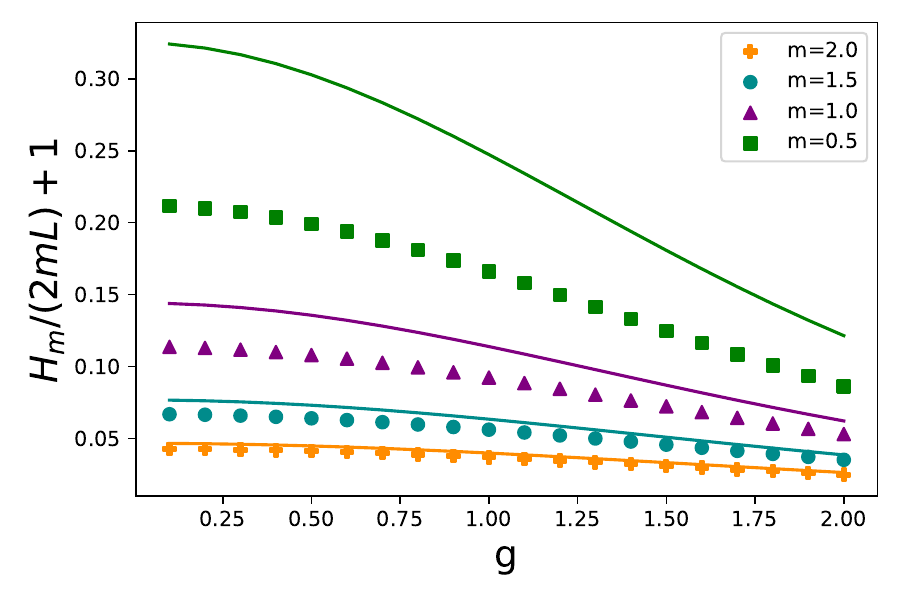}
\caption{Expectation of the mass term in the vacuum state as a function of $g$ for different quark masses on a lattice with $3$ physical sites. The solid lines were computed using the untruncated Hamiltonian in Eq.~\eqref{eq:QCD1DGauge} and the points were computed by projecting out all states with multiple (anti)quarks on a single site.}
    \label{fig:heavy_mass_exp}
\end{figure}

To explore what a heavy quark truncation looks like in practice, SU(2) lattice gauge theory in one dimension with one flavor of quark will be used as a test model. The Hamiltonian describing this theory with staggered fermions on $2L$ staggered sites is given by
\begin{align}
    & \hat{H} = \hat{H}_{Kin} + \hat{H}_m + \hat{H}_{E} \nonumber \\
    & \hat{H}_{Kin} = \sum_{x,a,b} \frac{1}{2} \hat{\psi}_{x,a}^\dagger \hat{U}^{a,b}_{x,x+1} \hat{\psi}_{x+1,b} + \text{h.c.} \nonumber \\
    & \hat{H}_m = m \sum_{x,a} (-1)^x \hat{\psi}_{x,a}^\dagger \hat{\psi}_{x,a} \nonumber \\
    & \hat{H}_E = \sum_{x,c} \frac{g^2}{2} \hat{E}_{x,x+1}^c\hat{E}_{x,x+1}^c \ \ \ .
    \label{eq:QCD1DGauge}
\end{align}
where $\hat{\psi}_{x,a}$ is the quark field at site $x$ with color $a$, $\hat{U}^{a,b}_{x,x+1}$ are components of parallel transporters between sites $x$ and $x+1$ and $\hat{E}_{x,x+1}^c$ is the corresponding electric field operator~\cite{kogut1975hamiltonian}. In the heavy quark limit, this Hamiltonian is dominated by $\hat{H}_m$. The quark fields can be truncated by restricting each (anti) quark site to have at most one (anti) quark. The coupling of these states to states with multiple quarks on a single site will be suppressed by factors of $1/m$. This truncation is analogous to the truncation of electric fields commonly done for quantum simulations of lattice gauge theories. By truncating local degrees of freedom by removing their highest energy components, the low energy states of the truncated theory will approximate the low energy states of the untruncated theory. The size of the deviations depends on the truncations as discussed in Appendix~\ref{sec:HamDeriv}. To understand the effect of this truncation, the vacuum of the Hamiltonian in Eq.~\eqref{eq:QCD1DGauge} can be compared to the vacuum where all states with multiple (anti)quarks on a single site are projected out. Fig.~\ref{fig:heavy_electric_exp} shows the electric energy for the two different truncations and Fig.~\ref{fig:heavy_mass_exp} shows the expectation of the mass term. The untruncated calculations were performed by integrating out the gauge field as in Ref.~\cite{Atas_2021}. Both of these figures show that at large $m$, this heavy quark truncation is in agreement with the untruncated theory. As $m$ is lowered, quantitative agreement is lost. However, one could use similarity renormalization group techniques to mitigate the effects of this truncation~\cite{ciavarella2023quantum}. With the application of these methods, one can improve the performance of this approach at smaller masses and approach closer to the continuum limit which occurs as $m\rightarrow0$. While this truncation may not be able to reach the continuum limit, it can be used to develop computational tools that can be applied in other formulations that can reach the continuum limit.

\begin{figure}
    \centering
    \includegraphics[width=8.6cm]{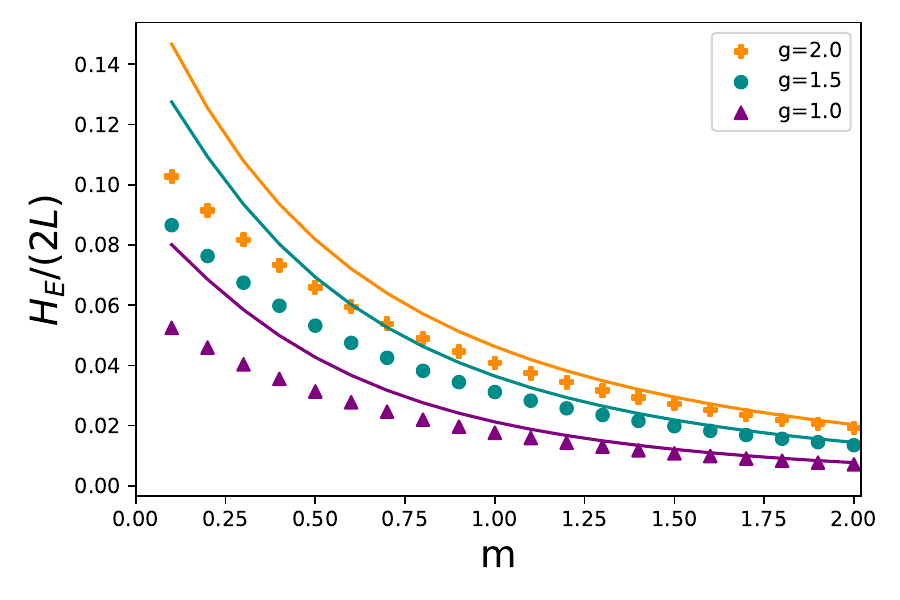}
\caption{Expectation of the electric energy in the vacuum state as a function of quark mass for a lattice with $4$ physical sites. The solid lines were computed using the untruncated Hamiltonian in Eq.~\eqref{eq:QCD1DGauge} and the points were computed using the Hamiltonian in Eq.~\eqref{eq:QCD1DQubit}.}
    \label{fig:electric_exp}
\end{figure}

\begin{figure}
    \centering
    \includegraphics[width=8.6cm]{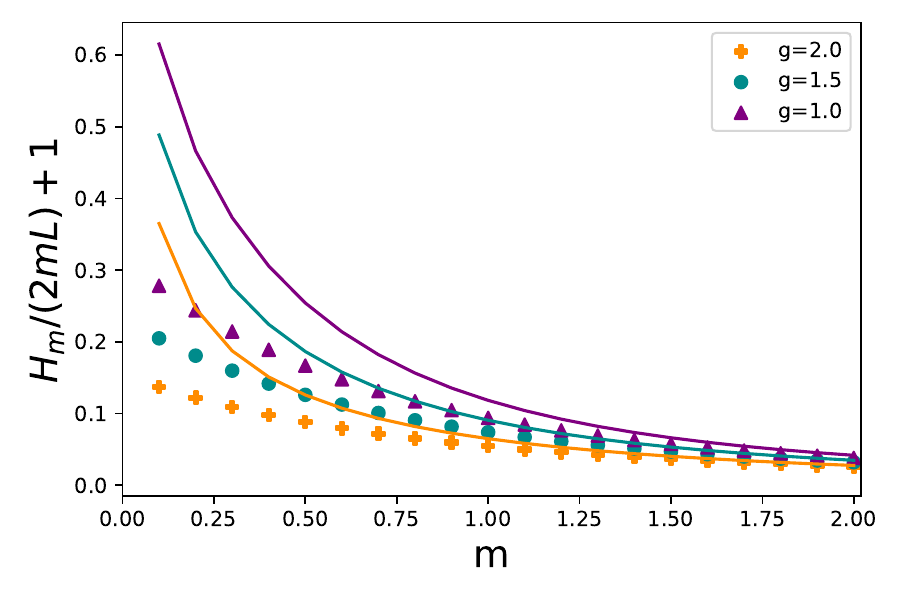}
\caption{Expectation of the mass term in the vacuum state as a function of quark mass for a lattice with $4$ physical sites. The solid lines were computed using the untruncated Hamiltonian in Eq.~\eqref{eq:QCD1DGauge} and the points were computed using the Hamiltonian in Eq.~\eqref{eq:QCD1DQubit}.}
    \label{fig:mass_exp}
\end{figure}

Performing a simulation of a gauge theory on a quantum computer requires a truncation of the theory to finite degrees of freedom. In one spatial dimension, this can be done by integrating out the gauge fields or truncating them. In higher spatial dimensions, it is necessary to truncate the gauge fields. This can be done in a gauge invariant manner by truncating in the electric basis. The performance of this truncation depends on the value of $g$. At large $g$, a harsh truncation can be used, but as $g$ is lowered more representations will need to be kept. In this work, the gauge fields will be truncated at $j=\frac{1}{2}$, which will allow mapping this theory directly onto qubits. At this truncation, the physical Hilbert space will be spanned by states filled with strings of electric flux stretching over links that connect quark anti-quark pairs. This Hilbert space can be represented on qubits by assigning a qubit to each link where the $0$ state corresponds to the link being unexcited and the $1$ state corresponds to the link being excited to $j=\frac{1}{2}$ state. Sites will have a quark present if only one of the neighboring links is excited. With this qubit basis, the Hamiltonian is
\begin{align}
    & \hat{H} = \hat{H}_{Kin} + \hat{H}_m + \hat{H}_{E} \nonumber \\
    & \hat{H}_{Kin} = \sum_{l} 
\frac{1}{\sqrt{2}}\hat{P}_{0,l} \hat{X}_{l+1} \hat{P}_{0,l+2} + \frac{1}{2\sqrt{2}}\hat{P}_{1,l} \hat{X}_{l+1} \hat{P}_{1,l+2} \nonumber \\
    & \hat{H}_m = m \sum_{l}  \hat{P}_{0,l} \hat{P}_{1,l+1} + \hat{P}_{1,l} \hat{P}_{0,l+1} \nonumber \\
    & \hat{H}_E = \sum_{l} \frac{3}{8} g^2 \hat{P}_{1,l} \ \ \ ,
    \label{eq:QCD1DQubit}
\end{align}
where $\hat{P}_{0,l}$ is the projector onto the $0$ state of qubit $l$, $\hat{P}_{1,l}$ is the projector onto the $1$ state of qubit $l$, and $\hat{X}_l$ is the Pauli $\hat{X}$ operator on qubit $l$. A detailed derivation of this Hamiltonian is in Appendix~\ref{sec:HamDeriv}. At this stage, it is worth commenting on what features of this truncated Hamiltonian came from having an SU(2) gauge symmetry in the untruncated theory. For a generic gauge group, one can integrate out the fermions and truncate at the lowest-lying electric fields. For gauge groups with complex representations (such as U(1) or SU(3)), one needs to specify an orientation on the links. For example, the representations of U(1) are specified by an integer $n$, and the lowest-lying representations have $n=0,\pm 1$. This requires a qutrit to be assigned to each link. Note that at $\theta=\pi$, a truncated $U(1)$ model can be mapped onto qubits~\cite{surace2020lattice} and one obtains a Hamiltonian similar to Eq.~\eqref{eq:QCD1DQubit} except without the $\hat{P}_{1,l} \hat{X}_{l+1} \hat{P}_{1,l+2}$ term. It is only for groups with real representations such as SU(2), that qubits can be used at this truncation at non-zero $\theta$. SU(2) is distinguished from these other groups by the numerical values of the coefficients in the truncated Hamiltonian. As a probe of this truncation, the expectation of the electric energy for the vacuum states of the Hamiltonian in Eq.~\eqref{eq:QCD1DQubit} and Eq.~\eqref{eq:QCD1DGauge} was computed at different values of $g$ and $m$ on a lattice with $4$ physical sites. The results are shown in Fig.~\ref{fig:electric_exp}. At large values of $g$ and $m$, this truncation works well as expected. Somewhat surprisingly, this truncation still holds at smaller $g$, provided that $m$ is kept large. The truncation works in this regime because states with larger electric fields also need to have more quarks present to satisfy Gauss's law. Fig.~\ref{fig:mass_exp} shows the vacuum expectation of the mass term in the Hamiltonian which displays similar behavior. 

\FloatBarrier
\section{State Preparation and Measurement}
Studying the real-time evolution of QCD on quantum computers requires the ability to prepare physically relevant states such as the vacuum or states containing hadrons. One also needs to be able to perform measurements at the end of the time evolution to determine how hadrons have propagated. It has previously been proposed to solve these problems by performing adiabatic switching from a free Hamiltonian at the beginning and end of the time evolution~\cite{jordan2011quantum,jordan7115quantum,preskill2018simulating,jordan2014quantum,jordan2012quantum}. However, performing adiabatic switching leads to deep quantum circuits that can be difficult to implement in practice. Variational algorithms to prepare the vacuum and single hadron states with shorter circuits have been developed and implemented on existing quantum computers~\cite{klco2018quantum,Atas_2021,muschik2023simulating,zhang2021simulating,ciavarella2022preparation,muschik2023simulating,farrell2023preparations2,farrell2023scalable,Farrell:2024fit}. The use of scalable circuits in these algorithms allows the variational circuits to be optimized using simulations of small lattices on classical computers and then implemented to prepare states on large lattices on quantum computers~\cite{farrell2023scalable,Farrell:2024fit,gustafson2024surrogate}. In addition to preparing the vacuum or known states, variational algorithms such as concurrent-VQE have been developed to prepare multiple low-lying eigenstates with the same circuit applied to different initial states~\cite{xu2023concurrent,guo2024concurrent,jaiswal2024simulating}. In this work, concurrent VQE will be combined with scalable circuits to develop a circuit that can be used to prepare the vacuum and single meson eigenstates from a product state and plane wave states, respectively. This procedure will be called Scalable Circuit Concurrent Variational Quantum Eigensolver (SC$^2$-VQE). These circuits then can be used to prepare states with isolated meson wavepackets depending on the initial state put into the circuit. Additionally, by applying the inverse of the state preparation circuit at the end of the simulation, it will be possible to perform measurements of hadron positions. This procedure is essentially the same as that proposed by Jordan, Lee, and Preskill, except the state preparation and measurement is done variationally instead of adiabatically~\cite{jordan2011quantum,jordan7115quantum,preskill2018simulating,jordan2014quantum,jordan2012quantum,Liu_2022}.

As an explicit example, this method will be applied to find the vacuum and single meson states of the Hamiltonian in Eq.~\eqref{eq:QCD1DQubit}. Instead of constructing a circuit with ADAPT-VQE at each lattice size, a fixed ansatz circuit $\hat{U}(\Vec{\theta})$ given by
\begin{align}
    \hat{U}(\Vec{\theta}) =& e^{i \sum_{x \text{ odd}} \theta_0 \hat{P}_{0, x-1} \hat{Y}_x \hat{P}_{0, x+1} + \theta_1 \hat{P}_{1, x-1} \hat{Y}_x \hat{P}_{1, x+1} } \nonumber \\ 
    & e^{i \sum_{x \text{ even}} \theta_0 \hat{P}_{0, x-1} \hat{Y}_x \hat{P}_{0, x+1} + \theta_1 \hat{P}_{1, x-1} \hat{Y}_x \hat{P}_{1, x+1} } \ \ \ ,
    \label{eq:vacCirc}
\end{align}
will be used. This circuit can be motivated by considering the time evolution unitary generated by adiabatic switching from the strong coupling limit. Eigenstates of Eq.~\eqref{eq:QCD1DQubit} can be prepared from eigenstates of the strong coupling limit by adiabatically turning on the hopping term. If this is done in an interaction picture where $\hat{H}_0 = \hat{H}_E + \hat{H}_m$, $\hat{V}(t) = e^{\epsilon t} \hat{H}_K$, the eigenstates of the full theory, $\ket{\psi_s}$ are related to the strong coupling eigenstates, $\ket{s}$, by 
\begin{equation}
    \ket{\psi_s} = \lim_{\epsilon\rightarrow0} \mathcal{T} e^{-i \int_{-\infty}^0 dt \ e^{\epsilon t} e^{i( \hat{H}_E + \hat{H}_m)t } \hat{H}_K e^{-i( \hat{H}_E + \hat{H}_m)t } } \ket{s} \ \ \ .
\end{equation}
Evaluating this expression to first order in the Magnus expansion yields 
\begin{align}
    & \ket{\psi_s}  \approx \nonumber \\
    & e^{i \left(\sum_x \frac{1}{\sqrt{2}} \frac{1}{3/8g^2 + 2m}  \hat{P}_{0, x-1} \hat{Y}_x \hat{P}_{0, x+1} + \frac{1}{2\sqrt{2}}\frac{1}{3/8g^2 - 2m}  \hat{P}_{1, x-1} \hat{Y}_x \hat{P}_{1, x+1} \right)} \ket{s} \ .
\end{align}
The proposed ansatz unitary is a Trotterized version of this unitary where the coefficients of the operators have been taken to be free parameters. Note that this ansatz is only expected to work well for couplings that result in short correlation lengths. A variational ansatz for systems with longer correlation lengths can be constructed by using more Trotter steps and higher-order terms in the Magnus expansion. Alternatively, a circuit ansatz could be built up iteratively through the use of ADAPT-VQE.

The parameters $\Vec{\theta}$ will be found through concurrent VQE. Concurrent VQE finds multiple eigenstates of a Hamiltonian by minimizing $\sum_i \bra{\psi_i} U^\dagger(\Vec{\theta}) \hat{H} U(\Vec{\theta})  \ket{\psi_i}$ where $\ket{\psi_i}$ are orthogonal initial states~\cite{xu2023concurrent}. The eigenstates found by concurrent VQE will be given by $U(\Vec{\theta})\ket{\psi_i}$. In this work, the vacuum states and single meson states will be prepared using concurrent VQE. The initial state for the vacuum will be given by the state with all qubits set to 0, $\ket{0}$. The initial state for the single meson eigenstates will be given by 
\begin{equation}
    \ket{p} = \frac{1}{\sqrt{2L}} \sum_x e^{i p x} \ket{x} \ \ \ ,
\end{equation}
where $p=0,\pm \frac{2\pi}{2L},\pm \frac{2\pi}{2L}\times2,\cdots,\frac{2\pi}{2L}\times L$ and $\ket{x}$ is a state with the qubit at site $x$ in the 1 state and all other qubits in the 0 state. In this work, all optimization will be performed using classical computers so instead of minimizing the energy, the average overlap with the eigenstates will be maximized. Explicitly, the quantity maximized in this work is
\begin{equation}
    O(\Vec{\theta}) = \frac{1}{2L+1}\left(\abs{\bra{\text{Vac}}U(\Vec{\theta})\ket{0}}^2 + \sum_p \abs{\bra{\psi_p}U(\Vec{\theta})\ket{p}}^2 \right) \ \ \ ,
\end{equation}
where $\ket{\text{Vac}}$ is the vacuum state of the full Hamiltonian and $\ket{\psi_p}$ is the single meson eigenstate with momentum $p$. Once $U(\Vec{\theta})$ has been optimized, it can be used to prepare meson wavepacket states by first preparing the state $\sum_p \phi(p) \ket{p}$ and then applying $U(\Vec{\theta})$ where $\phi(p)$ is the wavefunction specifying the shape of the wavepacket. In addition to state preparation, this procedure allows one to measure the location of mesons in a simulation. Explicitly, the overlap of a single meson wavepacket with wavefunction $\phi(p)$ with a generic state $\ket{\psi}$ is $\sum_p \phi^*(p) \bra{p}U^\dagger(\Vec{\theta})\ket{\psi}$. If $\phi(p)$ is chosen to be a delta function centered around lattice site $x$, this expression becomes $\bra{x}U^\dagger(\Vec{\theta})\ket{\psi}$. Therefore, by applying $U^\dagger(\Vec{\theta})$ before performing measurements, lines of qubits in the $\ket{1}$ state can be interpreted as having a meson present in that region, while qubits in the $\ket{0}$ state correspond to the vacuum. 

\begin{figure}
    \centering
    \includegraphics[width=8.6cm]{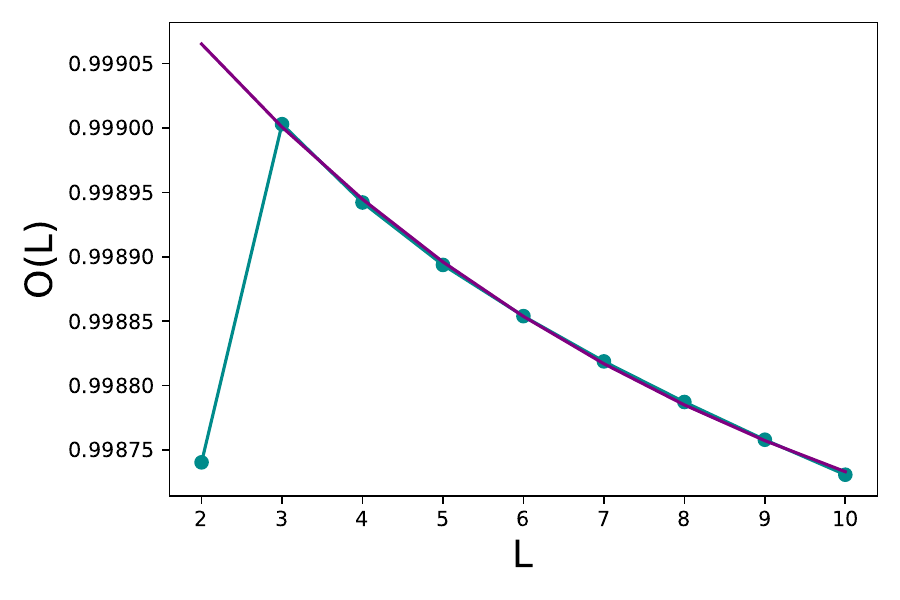}
    \caption{Average fidelity of the states prepared with concurrent VQE as a function of lattice size for $g=2$ and $m=1$. The orange line is an exponential fit excluding the first point.}
    \label{fig:AvgOverlap}
\end{figure}

\begin{figure}
    \centering
    \includegraphics[width=8.6cm]{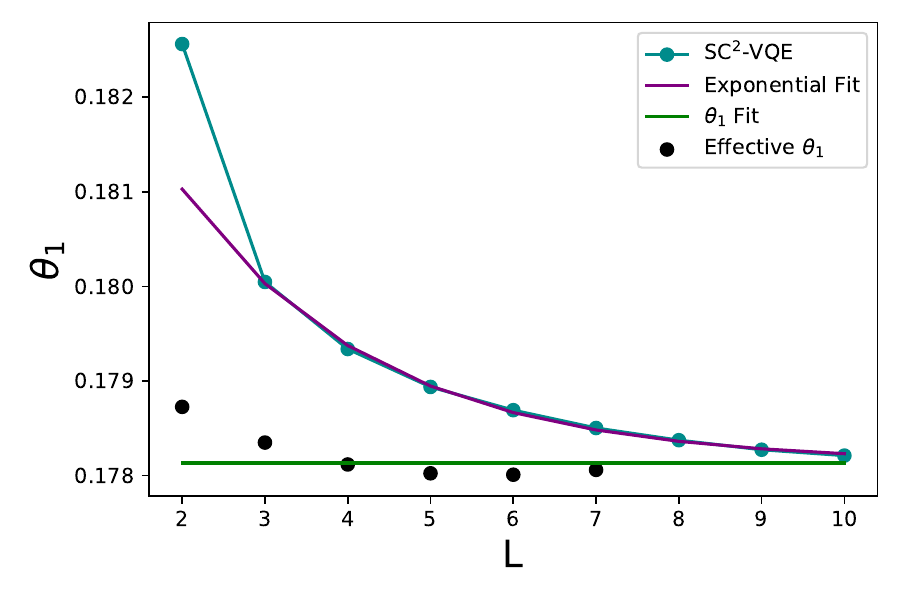}
    \caption{Optimal value of $\theta_1$ computed using concurrent VQE as a function of lattice size for $g=2$ and $m=1$.}
    \label{fig:theta1L}
\end{figure}

\begin{figure}
    \centering
    \includegraphics[width=8.6cm]{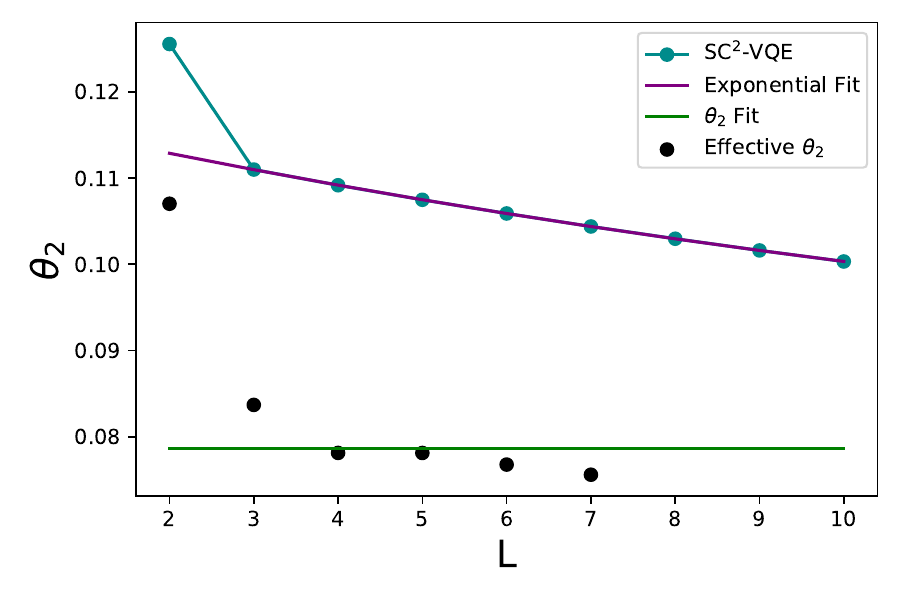}
    \caption{Optimal value of $\theta_2$ computed using concurrent VQE as a function of lattice size for $g=2$ and $m=1$.}
    \label{fig:theta2L}
\end{figure}

This procedure was performed for the Hamiltonian in Eq.~\eqref{eq:QCD1DQubit} with $g=2$ and $m=1$ and periodic boundary conditions. Fig.~\ref{fig:AvgOverlap} shows the optimal value of $O(\Vec{\theta})$ as a function of lattice size $L$, for $2\leq L \leq 10$. For $L>2$, the average overlap decreases exponentially with system size as would be expected. The optimal values of $\theta_1$ and $\theta_2$ are shown in Fig.~\ref{fig:theta1L} and Fig.~\ref{fig:theta2L} respectively. Both $\theta_i$ were fit to an exponential, excluding the first point. The $L\rightarrow\infty$ limit of both fits is in agreement with an effective infinite volume $\theta_i$ given by 
\begin{equation}
    \theta_{i, \ \text{eff.}} = \frac{\theta_i(L) \theta_i(L+3) - \theta_i(L+1) \theta_i(L+2)}{\theta_i(L+3)+\theta_i(L) - \theta_i(L+1)-\theta_i(L+2)} \ \ \ .
\end{equation}
This expression for $\theta_{i, \ \text{eff.}}$ was derived by assuming an exponential dependence on system size and solving to isolate the $L\rightarrow\infty$ as done in Ref.~\cite{farrell2023scalable}. This agreement suggests that the choice of an exponential fit function is correct for extrapolations to larger system sizes.

\FloatBarrier
\section{String Breaking}

In highly inelastic collisions, large numbers of $q\Bar{q}$ pairs are produced. At long times, $q\Bar{q}$ pairs will evolve to a superposition of different numbers of hadrons and these hadrons are what would ultimately be measured in a collider experiment. This is a real-time process that is difficult to simulate classically but may be accessible on quantum computers. Previous works have studied this process by performing simulations with dynamical light quarks and inserting external charges that correspond to moving heavy quarks~\cite{banerjee2012atomic,zhang2023observation,Florio_2023,lee2023liouvillian,florio2024quantumsimulation,farrell2024stepsquantum,de2024observationstring,cochran2024visualizing}. In the process of string breaking, there is a rapid growth in entanglement entropy followed by local observables thermalizing. Recent quantum simulations of a discrete gauge theory have also demonstrated a novel mechanism for string breaking beyond the traditional explanation in terms of the Schwinger effect~\cite{de2024observationstring}.

\subsection{Classical Simulation}
In this work, a state with a $q\Bar{q}$ pair will be prepared by flipping a line of qubits to the $\ket{1}$ state, i.e.
\begin{align}
    \ket{\psi_S} = U(\Vec{\theta})\prod_{x \in S} \hat{X}_x\ket{0} \ \ \ ,
    \label{eq:stringCirc}
\end{align}
where $S$ is the set of links in-between the sites where the quark and anti-quark have been placed and $U(\Vec{\theta})$ is the state preparation circuit determined using VQE. Note that long strings can be interpreted as excited states of a single meson. However, it should be emphasized that these states are not eigenstates of the Hamiltonian as the variational circuit has not been optimized to prepare excited states of mesons.

\begin{figure}
    \centering
    \includegraphics[width=8.6cm]{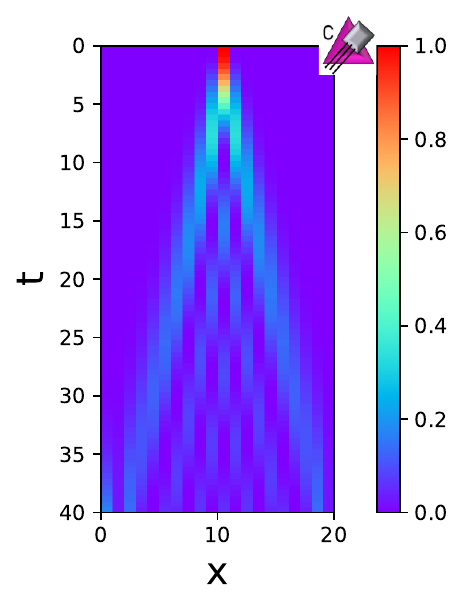}
    \caption{Evolution of $\hat{P}_{1,x}$ as a function of time and position for a string of length $1$ on a lattice with $20$ staggered sites and $g=2$, $m=1$.}
    \label{fig:string1P1}
\end{figure}

\begin{figure}
    \centering
    \includegraphics[width=8.6cm]{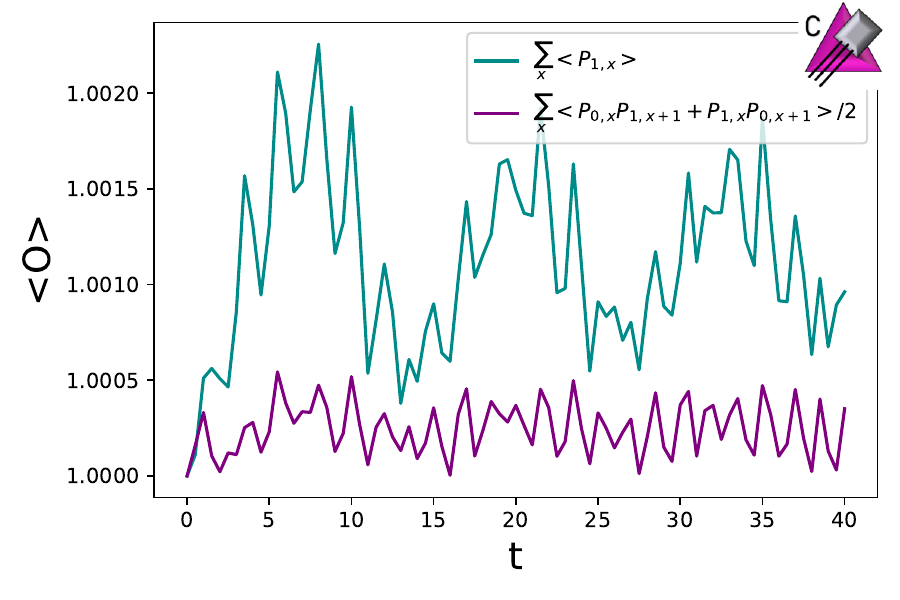}
    \caption{Evolution of $\sum_x \hat{P}_{1,x}$ and  and $\sum_x \hat{P}_{0,x} \hat{P}_{1,x+1} + \hat{P}_{1,x} \hat{P}_{0,x+1}$ as a function of time for a string of length $1$ on a lattice with $20$ staggered sites and $g=2$, $m=1$.}
    \label{fig:string1TotalP1}
\end{figure}

When only one qubit is flipped, this is a single meson wavefunction. Therefore, one would expect to not observe any string breaking for a string of this size. The dynamics of the system will be studied using the evolution of $\hat{P}_{1,x}$ and $\hat{P}_{0,x} \hat{P}_{1,x+1} + \hat{P}_{1,x} \hat{P}_{0,x+1}$. Following the discussion in the previous section, after undoing $U(\Vec{\theta})$, a qubit being in the $1$ state with its neighbors in the $0$ state corresponds to a meson being present at that qubit. Therefore, $(\hat{P}_{0,x} \hat{P}_{1,x+1} + \hat{P}_{1,x} \hat{P}_{0,x+1})/2$ can be used to track the position and number of mesons in the simulation. The motivation for this operator can be seen in the strong coupling limit where it indicates the presence of a quark or anti-quark at the site between links $x$ and $x+1$. Note that $\hat{P}_{1,x}$ does not work for this purpose, as it is insensitive to the state of the neighboring qubits. This is an issue when there are multiple mesons present in the system. Fig.~\ref{fig:string1P1} shows the evolution of $\hat{P}_{1,x}$ as a function of time and position for a single link string on a lattice with $20$ staggered sites and $g=2$, $m=1$. This figure shows an excitation at the center of the lattice propagating out in both directions. The shape of the propagation is consistent with either a single meson wavepacket that has support on both positive and negative momentum modes or a string breaking into two mesons that propagate in different directions. To distinguish between these two cases, one can compute the expectation values of $\sum_x \hat{P}_{1,x}$ and $\sum_x \hat{P}_{0,x} \hat{P}_{1,x+1} + \hat{P}_{1,x} \hat{P}_{0,x+1}$. The expectation of these observables is shown in Fig.~\ref{fig:string1TotalP1}. This figure shows that $\sum_x \hat{P}_{1,x}$ fluctuates around one and $\sum_x \hat{P}_{0,x} \hat{P}_{1,x+1} + \hat{P}_{1,x} \hat{P}_{0,x+1}$ fluctuates around two. This indicates that there is only a single meson present in the system and the string is not breaking. These small fluctuations are likely due to imperfections in the state preparation circuit.

\begin{figure}
    \centering
    \includegraphics[width=8.6cm]{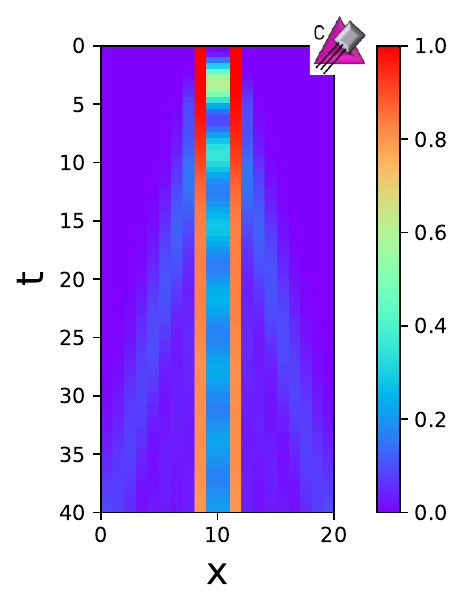}
    \includegraphics[width=8.6cm]{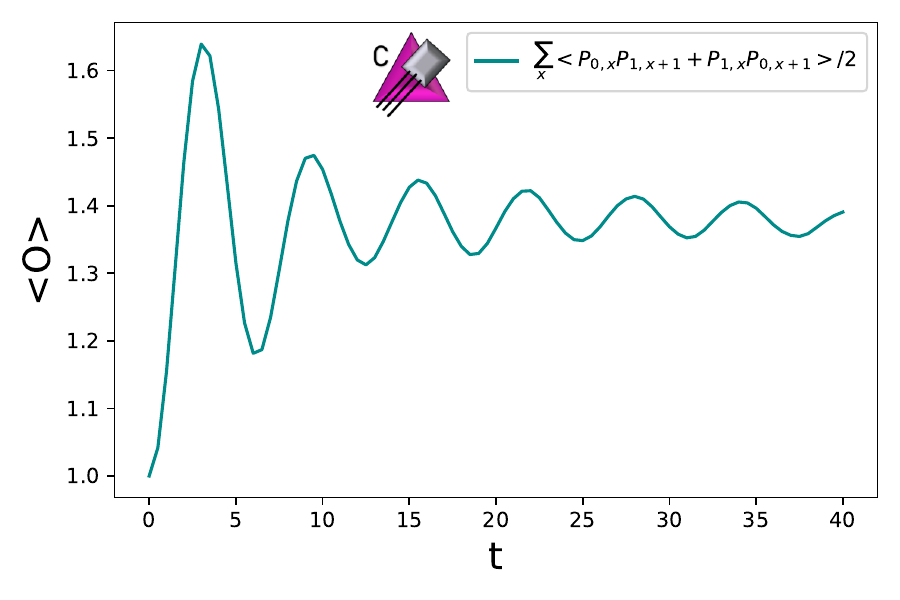}
    \caption{Evolution of $\hat{P}_{0,x} \hat{P}_{1,x+1} + \hat{P}_{1,x} \hat{P}_{0,x+1}$ as a function of time and position for a string of length $3$ on a lattice with $20$ staggered sites and $g=2$, $m=1$.}
    \label{fig:string3P1}
\end{figure}

To observe string breaking, it is necessary to prepare a longer string on the lattice. Fig.~\ref{fig:string3P1} shows the evolution of $\hat{P}_{0,x} \hat{P}_{1,x+1} + \hat{P}_{1,x} \hat{P}_{0,x+1}$ for a string that extends over $3$ staggered sites. As this figure shows, the evolution begins with a single $q\Bar{q}$ pair and as the system evolves, additional $q\Bar{q}$ pairs are created and break the string. At late times, the system settles into a superposition of a single meson state and multi-meson states. Near the original location of the string, $\hat{P}_{0,x} \hat{P}_{1,x+1} + \hat{P}_{1,x} \hat{P}_{0,x+1}$ settles into oscillations around a constant value, reminiscent of the thermalization observed in Ref~\cite{florio2024quantumsimulation}. The relatively slow dynamics in the center of the string can be understood as a suppression of the Schwinger mechanism by the heavy quark mass~\cite{Lerose:2019jrs,Verdel:2019chj}.

\begin{figure}
    \centering
    \includegraphics[width=8.6cm]{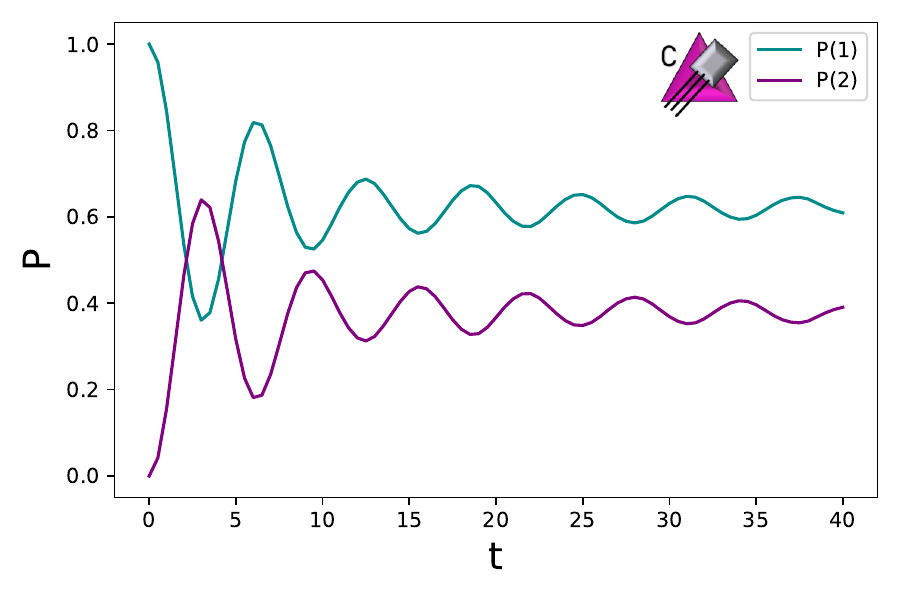}
    \caption{Probability distribution for the number of mesons as a function of time, beginning with a length $3$ string on a lattice with $20$ staggered sites and $g=2$, $m=1$. }
    \label{fig:ProbMeson}
\end{figure}

The average number of mesons in the system is determined by measuring $\hat{O} = \frac{1}{2}\sum_x\hat{P}_{0,x} \hat{P}_{1,x+1} + \hat{P}_{1,x} \hat{P}_{0,x+1}$. The total number of mesons in a given basis state is determined by the number of neighboring qubits in different states (i.e. how many times the string $01$ or $10$ occurs in the bit string defining the basis state). This can be measured directly on a quantum computer to determine $P(n)$, the probability that the system contains $n$ mesons. However, the total number of mesons in a state is a global observable which is highly sensitive to noise. An alternative approach is to measure local observables, for which effective error mitigation techniques exist~\cite{trexmit,urbanek2021mitigating,Rahman:2022rlg,rahman2022real,farrell2023scalable,ciavarella2024quantum,urbanek2021mitigating,he2020zero,pascuzzi2022computationally,Farrell:2024fit}, and determine $P(n)$ from there. Explicitly, one can measure and apply error mitigation techniques to $\bra{\psi} \hat{O}^k \ket{\psi}$ for some powers $k$. The system of equations
\begin{align}
    1 & = \sum_{n=1}^{N_{max}} P(n) \nonumber \\
    \bra{\psi} \hat{O}^k \ket{\psi} & = \sum_{n=1}^{N_{max}} n^k P(n) \ \ \ ,
\end{align}
can be set up and solved for $P(n)$, assuming some maximum number of mesons $N_{max}$ in the state. $N_{max}$ can be determined using the energy of the state and the mass of the meson. For the length $3$ string simulated above, the energy of the state is between $2m$ and $3m$ where $m$ is the mass of the meson. Fig.~\ref{fig:ProbMeson} shows $P(1)$ and $P(2)$ for the length $3$ string as a function of time. At short times, there are large oscillations in the probabilities, and at late times there are small oscillations around a long time limit. Previous work on confinement in spin models has observed similar oscillations and related them to the quasiparticle spectrum of the model~\cite{Verdel:2019chj,Lerose:2019jrs,Lin:2016egw,Liu:2018fza,Kormos:2016osj}. It was shown in these models that these oscillations should decay at late times on large lattices~\cite{Lin:2016egw}. Similar behavior is expected in this theory.

\subsection{Quantum Simulation}

\begin{figure}
    \centering
    \includegraphics[width=8.6cm]{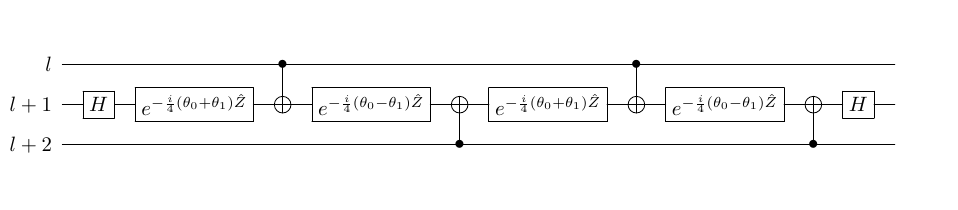}
    \includegraphics[width=8.6cm]{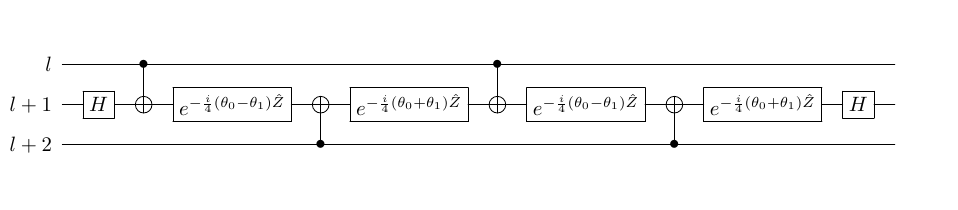}
    \caption{Two possible circuits that can be used to implement $e^{-i \theta_0 \hat{P}_{0,l} \hat{X}_{l+1} \hat{P}_{0,l+2} + \theta_1 \hat{P}_{1,l} \hat{X}_{l+1} \hat{P}_{1,l+2}}$}
    \label{fig:PXPCircuits}
\end{figure}

To perform a simulation of these dynamics on a quantum computer, the time evolution operator needs to be implemented. In this work, this will be done using a second-order Trotter formula. Explicitly, the time evolution operator will be approximated by
\begin{align}
    U_{Tr}(\Delta t) = & e^{-i \Delta t /2 \hat{H}_{K,o}} e^{-i \Delta t /2 \hat{H}_{K,e}} e^{-i \Delta t (\hat{H}_E + \hat{H}_m)} \nonumber \\
    & e^{-i \Delta t /2 \hat{H}_{K,e}} e^{-i \Delta t /2 \hat{H}_{K,o}} \nonumber \\
    \hat{H}_{K,e} = & \sum_{l\text{ even}} 
\frac{1}{\sqrt{2}}\hat{P}_{0,l} \hat{X}_{l+1} \hat{P}_{0,l+2} + \frac{1}{2\sqrt{2}}\hat{P}_{1,l} \hat{X}_{l+1} \hat{P}_{1,l+2} \nonumber \\
\hat{H}_{K,o} = & \sum_{l\text{ odd}} 
\frac{1}{\sqrt{2}}\hat{P}_{0,l} \hat{X}_{l+1} \hat{P}_{0,l+2} + \frac{1}{2\sqrt{2}}\hat{P}_{1,l} \hat{X}_{l+1} \hat{P}_{1,l+2}
\end{align}
The operator $e^{-i \Delta t (\hat{H}_E + \hat{H}_m)}$ can be decomposed into a sum of Pauli $\hat{Z}$ operators and products of $\hat{Z}$ operators on neighboring qubits. A circuit (with CNOT depth $4$) to implement this operator can be formed using standard textbook techniques~\cite{nielsen2001quantum}. Either of the two circuits in Fig.~\ref{fig:PXPCircuits} can be used to implement $e^{-i (\theta_0 \hat{P}_{0,l} \hat{X}_{l+1} \hat{P}_{0,l+2} + \theta_1 \hat{P}_{1,l} \hat{X}_{l+1} \hat{P}_{1,l+2})}$. The depth of the circuit can be reduced by using the upper circuit to evolve the even sublattice and then using the lower circuit to evolve the odd sublattice. The direction of the first CNOT in the lower circuit can be switched by conjugating it with Hadamard gates. This allows it to cancel the last CNOT gate in the upper circuit. Therefore, $e^{-i \Delta t /2 \hat{H}_{K,o}} e^{-i \Delta t /2 \hat{H}_{K,e}}$ can be implemented with a circuit that has a CNOT depth of $6$. Note that the same circuit design can be used to implement the state preparation unitary in Eq.~\eqref{eq:vacCirc} by conjugating the circuit with $\hat{S}$ gates.

\begin{figure}
    \centering
    \includegraphics[width=8.6cm]{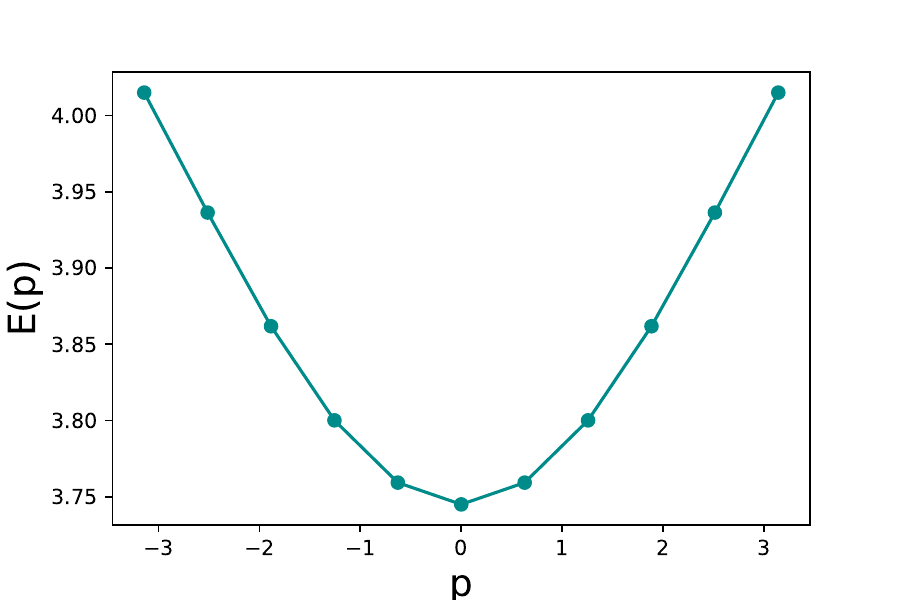}
    \caption{Dispersion relation for mesons on a lattice with $10$ physical sites and $g=2$, $m=1$.}
    \label{fig:MesonDispersion}
\end{figure}

Once the time evolution circuit has been chosen, a step size $\Delta t$ must be chosen. Previous work has suggested that Trotterization can be understood as introducing a temporal lattice with spacing $\Delta t$~\cite{carena2021lattice}. With this perspective, there is no benefit to taking $\Delta t$ to be significantly smaller than the spatial lattice spacing. To compare these spacings, one needs to determine the lattice anisotropy, or equivalently the speed of light on the lattice. This can be done using the dispersion relation of the mesons. At low energies, the dispersion relation is expected to take the form
\begin{equation}
    E^2(p) = E^2(0) + p^2 c^2 \ \ \ ,
\end{equation}
where $c$ is the speed of light on the lattice and $p$ is the momentum of the meson. The momentum of a meson state can be determined from the phase obtained by applying a translation by a single physical site (i.e. two staggered sites). The dispersion relation for a meson with $g=2$, $m=1$ on a lattice with $10$ physical sites is shown in Fig.~\ref{fig:MesonDispersion}. From this dispersion relation, we find $c\approx0.521$. 
To ensure Trotterization is not the dominant source of error in this calculation, the value $\Delta t = 0.5$ will be used. This corresponds to a temporal lattice spacing that is roughly one-quarter of the spatial lattice spacing.

\begin{figure}
    \centering
    \includegraphics[width=8.6cm]{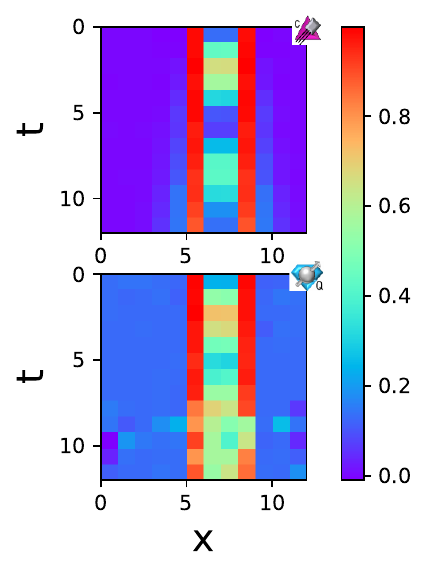}
    \includegraphics[width=8.6cm]{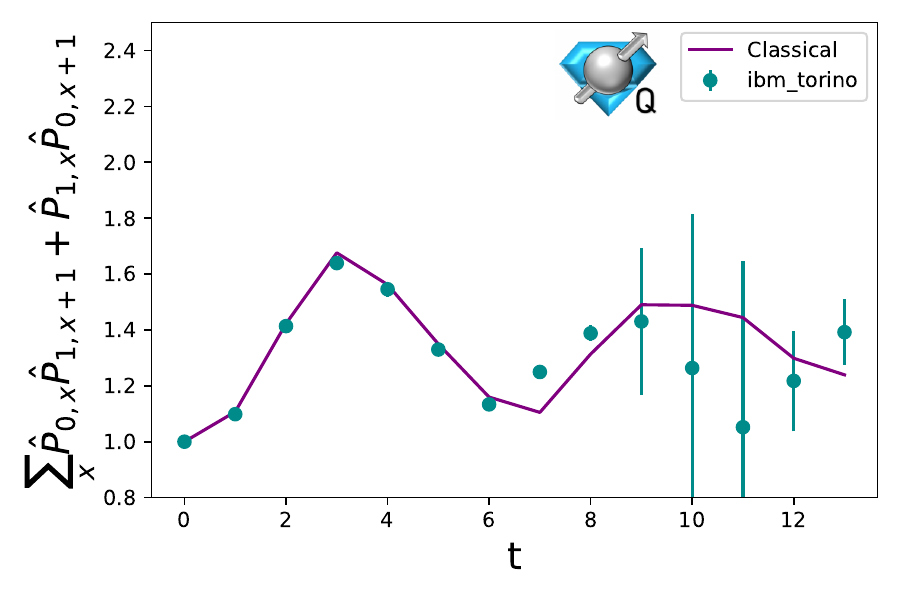}
    \caption{Evolution of $\hat{P}_{0,x} \hat{P}_{1,x+1} + \hat{P}_{1,x} \hat{P}_{0,x+1}$ as a function of time and position on a lattice with 12 staggered sites. The upper plot shows the classical simulation and the center plot shows the error mitigated results from {\tt ibm\_torino}. The lower plot shows the expectation of $\sum_x \hat{P}_{0,x} \hat{P}_{1,x+1} + \hat{P}_{1,x} \hat{P}_{0,x+1}$ as a function of time. All error bars were computed using a statistical bootstrap.}
    \label{fig:quarksIBM12}
\end{figure}

Quantum simulations of the truncated Hamiltonian in Eq.~\eqref{eq:QCD1DQubit} with $g=2$, $m=1$, and PBC were performed using the {\tt ibm\_torino} quantum processor~\cite{aleksandrowicz2019qiskit,ibmTorino}. To determine the circuit depth that can be successfully probed, a simulation was performed using $12$ qubits. A length 3 string was prepared in the center of the lattice and evolved with second-order Trotter steps with $\Delta t = 0.5$. The vacuum state was also prepared and evolved under the same time evolution circuit. Dynamical decoupling with the $XY4$ sequence was used to suppress the effects of noise~\cite{viola1999dynamical}. Additionally, the assignment of each qubit onto lattice sites was randomized for each circuit sent to IBM's hardware (while maintaining the required nearest neighbor connectivity). Pauli twirling was used to convert all noise into Pauli channels~\cite{urbanek2021mitigating,Rahman:2022rlg,rahman2022real} and TREX was used to correct measurement errors~\cite{trexmit}. Each time slice contains $80$ twirls and each twirl was sampled $8000$ times. Remaining errors were mitigated by using Operator Decoherence Renormalization (ODR) with the reference state being obtained by performing backward time evolution for the first half of the Trotter steps and forward time evolution for the second half~\cite{farrell2023scalable}. Up to $26$ Trotter steps were performed, reaching a CNOT depth of 428. Fig.~\ref{fig:quarksIBM12} shows the expectation of $\hat{P}_{0,x} \hat{P}_{1,x+1} + \hat{P}_{1,x} \hat{P}_{0,x+1}$ for each site as a function of time. For each site, the expectation of the vacuum's evolution was subtracted from that with the string present. With error mitigation, the quantum computer is able to accurately simulate the dynamics of the string breaking up to around $t=6$, corresponding to a CNOT depth of $204$. This is sufficient to see the initial oscillations in the number of mesons present in the system, but not to extract any observables at asymptotically long times. Up until around $t=13$ (corresponding to a CNOT depth of 428), some qualitative features of the time evolution remain, but quantitative accuracy is lost. Probing further times accurately would require either improved error mitigation or lower error rates on the quantum computer.

\begin{figure}
    \centering
    \includegraphics[width=8.6cm]{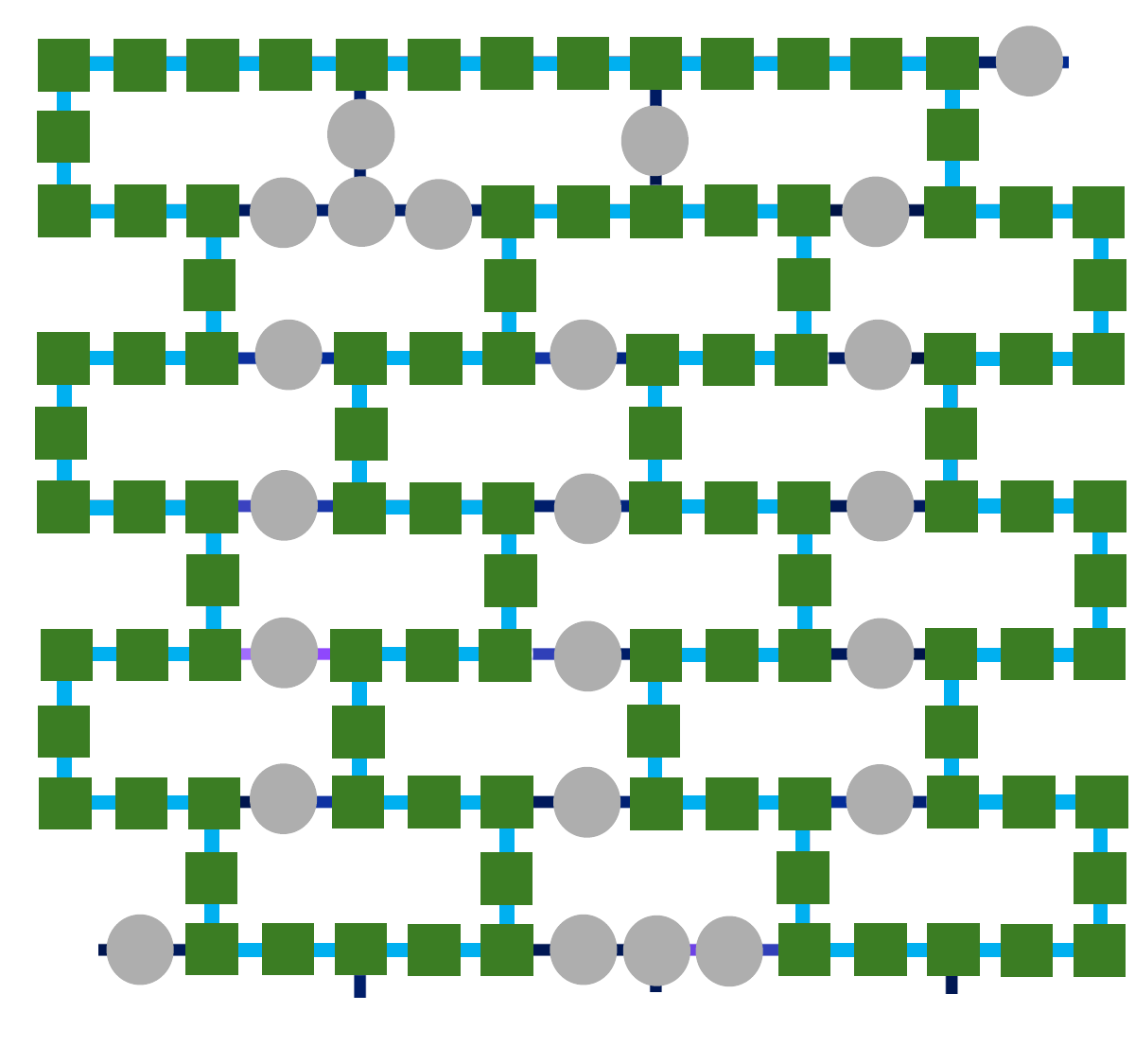}
    \caption{Qubits on {\tt ibm\_torino} used to simulate the lattice with $104$ staggered sites. The green squares connected by a light blue line represent the qubits used in the simulation. The light gray qubits were not used.}
    \label{fig:QubitAssignment}
\end{figure}

\begin{figure}
    \centering
    \includegraphics[width=9cm]{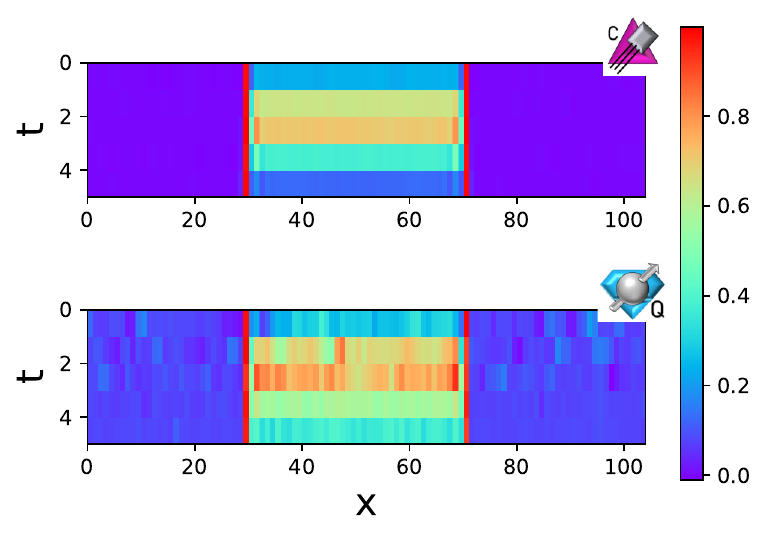}
    \includegraphics[width=8.6cm]{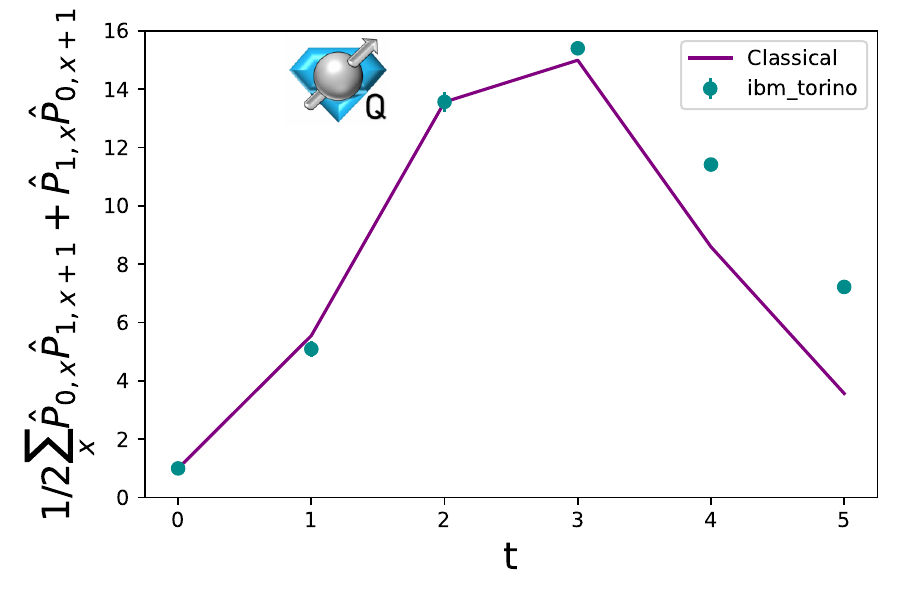}
    \caption{Evolution of $\hat{P}_{0,x} \hat{P}_{1,x+1} + \hat{P}_{1,x} \hat{P}_{0,x+1}$ as a function of time and position for a lattice with $104$ staggered sites. The upper plot shows the classical simulation and the center plot shows the error mitigated results from {\tt ibm\_torino}. The lower plot shows the expectation of $\sum_x \hat{P}_{0,x} \hat{P}_{1,x+1} + \hat{P}_{1,x} \hat{P}_{0,x+1}$ as a function of time.}
    \label{fig:quarksIBM104}
\end{figure}

A more clear demonstration of string breaking can be seen using a longer string on a larger lattice. With the selection of qubits shown in Fig.~\ref{fig:QubitAssignment}, it is possible to encode a lattice with 104 staggered sites with periodic boundary conditions onto {\tt ibm\_torino}. A string of length $41$ links was prepared and evolved using the same Trotterized circuit as discussed above. The angles in the vacuum preparation circuit were determined using the extrapolation discussed previously. The evolution was performed with up to ten Trotter steps resulting in a maximum CNOT depth of $172$ with a total of $8944$ CNOT gates. The results of the simulation are shown in Fig.~\ref{fig:quarksIBM104}. As in the $12$ qubit case, the initial oscillations are accurately reproduced on the quantum computer, but it is not possible to evolve to asymptotically long times and extract information about the number of mesons present at late times. Reaching these times will likely require improvements in both hardware error rates and mitigation strategies.

\section{Discussion}

In this work, a truncation of lattice gauge theories based on the heavy quark limit has been introduced. This truncation preserves gauge invariance and all other symmetries of the theory while simplifying the mapping onto qubits. This will enable near-term simulation of lattice gauge theories in this limit. The scalable VQE approach to state preparation has been generalized to prepare arbitrary single meson wavepackets. This extension also enables the measurement of the positions of mesons at later times by undoing the state preparation circuits before performing a measurement. This new formalism was used to perform classical simulations of string breaking in a $1+1D$ $SU(2)$ LGT. Performing state preparation and measurement with SC$^2$-VQE enabled a determination of the probability distribution for the number of mesons produced in this string-breaking process. Similar calculations were performed using {\tt ibm\_torino}, however, the accuracy of the simulation was lost before asymptotically long times could be probed.

The SC$^2$-VQE method introduced in this work should be generalizable to theories with other gauge groups and higher dimensions. Similar to other scalable circuits, this technique only relies on translation invariance and exponentially decaying correlations for low energy states~\cite{farrell2023scalable,Farrell:2024fit,gustafson2024surrogate}. For theories with multiple different propagating degrees of freedom (i.e. baryons and mesons), SC$^2$-VQE can be used by including the relevant initial states in the concurrent VQE. Additionally, it can be combined with ADAPT-VQE to iteratively build up ansatz circuits for more general systems. This will enable simulations of scattering between different types of hadrons and more detailed examinations of resulting final states. 

\begin{acknowledgements}
The authors would like to thank Zhiyao Li for many useful discussions. The authors would also like to thank Nikita Zemlevskiy, Marc Illa, Roland Farrell, Martin Savage, and Christian Bauer for useful discussions. This material is based upon work supported by the U.S. Department of Energy, Office of Science, National Quantum Information Science Research Centers, Quantum Systems Accelerator. Additional support is acknowledged from the U.S. Department of Energy (DOE), Office of Science under contract DE-AC02-05CH11231, partially through Quantum Information Science Enabled Discovery (QuantISED) for High Energy Physics (KA2401032). This research used resources of the Oak Ridge Leadership Computing Facility (OLCF), which is a DOE Office of Science User Facility supported under Contract DE-AC05-00OR22725. We acknowledge the use of IBM Quantum services for this work. The views expressed are those of the authors and do not reflect the official policy or position of IBM or the IBM Quantum team.
\end{acknowledgements}

\bibliographystyle{apsrev4-1}
\bibliography{ref}
\appendix
\onecolumngrid

\section{Derivation of Heavy Quark Hamiltonian}
\label{sec:HamDeriv}
The Hilbert space for SU(2) lattice gauge theory in $1+1$D is spanned by basis states of the form
\begin{equation}
    \ket{\psi} = \prod_l \ket{n_l,a_l}_l \ket{R_l,r_{l,L},r_{l,R}}_l \ \ \ ,
\end{equation}
where $\ket{n_l,a_l}_l$ is a basis state for the fermions at site $l$ and $\ket{R_l,r_{l,L},r_{l,R}}_l$ is a basis state for the gauge field on the link to the right of site $l$. For the fermions, $n_l$ specifies the number of fermions at the site. Since the fermions anti-commute, $n_l$ also specifies what irrep, $j$, of the gauge group the fermions are in on site $l$. $a_l$ denotes what component of the irrep the fermions are in. For $n_l=0$ or $n_l=2$, the fermions must be in a $j=0$ irrep and $a_l=0$. If $n_l=1$, then $j=\frac{1}{2}$, and $a_l=\pm\frac{1}{2}$. For the gauge fields, $R_l$ denotes a representation of SU(2), $r_{l,L}$ is a component of $R_l$ associated with the left side of the link and $r_{l,R}$ is a component of $R_l$ associated with the right side of the link. Gauge invariance requires that at each site the fermions and gauge fields from the neighboring links combine together to form a singlet under SU(2). With the convention that the first staggered site is $l=0$\footnote{This is necessary for determining the phase of the staggered mass term.}, the vacuum state of the electric and mass terms of the Hamiltonian has all even sites empty and all odd sites full, i.e.
\begin{equation}
    \ket{0} = \ket{0,0} \ket{0,0,0} \ket{2,0} \ket{0,0,0} \cdots \ \ \ .
\end{equation}
The heavy quark truncation introduced in this work removes basis states where the even sites have $2$ fermions present and the odd sites have $0$ fermions present. With the truncation of the gauge field at $j=\frac{1}{2}$, one can use the gauge constraint to integrate out the fermions and the left and right indices of the gauge fields. Explicitly due to the heavy quark truncation, if the links at $l-1$ and $l$ both have $j=0$ or $j=\frac{1}{2}$, the state of the fermions at site $l$ is fixed to be either $n_l=0$, if $l$ is even, or $n_l=2$, if $l$ is odd. If only one of links $l-1$ or link $l$ is excited, then $n_l=1$ and the irrep indices of the fermion must be contracted with those of the electric field. This allows us to use an assignment of qubits to each link to represent states at this electric and heavy quark truncation. The $\ket{0}$ state of the qubit corresponds to the link having $R_l=0$ and the $\ket{1}$ state of the qubit corresponds to the link having $R_l=\frac{1}{2}$. For electric basis states, the electric energy of a link is given by $\frac{g^2}{2}$ times the Casimir of the irrep on the link so for this encoding onto qubits the electric energy is given by
\begin{equation}
    \hat{H}_E = \sum_{l} \frac{3}{8} g^2 \hat{P}_{1,l} \ \ \ ,
\end{equation}
where $\hat{P}_{1,l}$ is the projection onto the $\ket{1}$ state for the qubit on link $l$. Up to an overall constant, the mass term of the Hamiltonian will be given by 
\begin{equation}
    \hat{H}_m = m \sum_{l}  \hat{P}_{0,l} \hat{P}_{1,l+1} + \hat{P}_{1,l} \hat{P}_{0,l+1} \ \ .
\end{equation}
The kinetic term of the Hamiltonian can be determined by considering a lattice with $4$ staggered sites and open boundary conditions. In the qubit basis, the electric vacuum state is $\ket{000}$. Denoting the hopping term between the second and third site by
\begin{equation}
    \hat{H}_2 = \frac{1}{2} \sum_{a,b} \left( \hat{\psi}_{1,a}^\dagger \hat{U}^{a,b}_{1,2} \hat{\psi}_{2,b} + \text{h.c.} \right) \ \ \ ,
\end{equation}
it can be seen that
\begin{align}
    \hat{H}_2 \ket{000} &= \frac{1}{\sqrt{2}} \ket{010} \nonumber \\
    \hat{H}_2 \ket{110} &= 0 \nonumber \\
    \hat{H}_2 \ket{011} &= 0 \nonumber \\
    \hat{H}_2 \ket{101} &= \frac{1}{2\sqrt{2}} \ket{111}
\end{align}
at this truncation. Therefore, in the qubit basis,
\begin{equation}
    \hat{H}_2 = \frac{1}{\sqrt{2}}\hat{P}_{0,0} \hat{X}_{1} \hat{P}_{0,2} + \frac{1}{2\sqrt{2}}\hat{P}_{1,0} \hat{X}_{1} \hat{P}_{1,2} \ \ \ .
\end{equation}
For a generic sized lattice, it follows that 
\begin{equation}
    \hat{H}_K = \frac{1}{\sqrt{2}}\hat{P}_{0,l} \hat{X}_{l+1} \hat{P}_{0,l+2} + \frac{1}{2\sqrt{2}}\hat{P}_{1,l} \hat{X}_{l+1} \hat{P}_{1,l+2} \ \ \ .
\end{equation}

To use this truncation in practice, one needs to understand the size of the errors induced by truncating the Hilbert space. The effects of truncating the electric field are well understood and there is an exponential convergence as the truncation is raised~\cite{tong2022provably}. The size of the errors from the heavy quark truncation can be estimated using traditional perturbation theory. Explicitly, one can write the Hamiltonian of the untruncated theory as $\hat{H} = \hat{H}_0 + \hat{V}$ where $\hat{H}_0$ is the full electric and mass term of the Hamiltonian with the kinetic term restricted to the states kept in the truncation. $\hat{V}$ contains all couplings to the states removed by the truncation of the Hilbert space. To leading order, the eigenstates $\ket{\psi}$ and energies $E$ of the untruncated Hamiltonian ($\ket{\psi_T}$ and $E_T$) by
\begin{align}
    E &= E_T + \bra{\psi_T} \hat{V} \ket{\psi_T} + \bra{\psi_T} \hat{V} \frac{1}{E_T - \hat{H}_0} \hat{V} \ket{\psi_T} \nonumber \\
    \ket{\psi} & = \ket{\psi}_T + \frac{1}{E_T - \hat{H}_0} \hat{V} \ket{\psi_T} \ \ \ .
\end{align}
With this truncation strategy, we have $\bra{\psi_T} \hat{V} \ket{\psi_T}=0$. Therefore, the suppression of the correction terms for the eigenstates and energies will come from the factor of $\frac{1}{E_T - \hat{H}_0}$. When $\hat{V}$ is applied to a truncated eigenstate, the resulting state will have either a link with an electric field above the cutoff $j_{max}$ (which makes $\frac{1}{E_T - \hat{H}_0} = \mathcal{O}(\frac{1}{g^2 j_{max}})$) or a pair of quarks on a site (which makes $\frac{1}{E_T - \hat{H}_0} = \mathcal{O}(\frac{1}{m})$). Therefore, all differences due to the electric and heavy quark truncations are either $\mathcal{O}(\frac{1}{g^2 j_{max}})$ or $\mathcal{O}(\frac{1}{m})$.

\FloatBarrier
\section{Tables}
The following tables contain the data obtained from {\tt ibm\_torino}. Note that the expectations listed here are given by the difference between the expectation for the state with a string present and the vacuum state evolved in time using Trotterization.

\begin{table}[]
\centering
\begin{tabular}{|c|c|c|}
\hline
$x$ & Classical Simulation & {\tt ibm\_torino} \\ 
\hline
0 & 0.0001 & 0.004$ \pm $0.004 \\ 
\hline
1 & -0.0004 & 0.014$ \pm $0.004 \\ 
\hline
2 & -0.0017 & 0.011$ \pm $0.005 \\ 
\hline
3 & -0.0059 & 0.002$ \pm $0.005 \\ 
\hline
4 & -0.0096 & -0.007$ \pm $0.005 \\ 
\hline
5 & 0.9861 & 0.98$ \pm $0.004 \\ 
\hline
6 & 0.1358 & 0.124$ \pm $0.005 \\ 
\hline
7 & 0.1347 & 0.121$ \pm $0.006 \\ 
\hline
8 & 0.9821 & 0.976$ \pm $0.004 \\ 
\hline
9 & -0.0085 & -0.016$ \pm $0.004 \\ 
\hline
10 & 0.0004 & -0.01$ \pm $0.005 \\ 
\hline
11 & 0.0011 & -0.003$ \pm $0.004 \\ 
\hline
\end{tabular}
\caption{The expectation of $\hat{P}_{0,x} \hat{P}_{1,x+1} + \hat{P}_{1,x} \hat{P}_{0,x+1}$ for $t=1.0$ displayed in Fig.\ref{fig:quarksIBM12}.}
\end{table}\begin{table}[]
\centering
\begin{tabular}{|c|c|c|}
\hline
$x$ & Classical Simulation & {\tt ibm\_torino} \\ 
\hline
0 & -0.0001 & 0.001$ \pm $0.006 \\ 
\hline
1 & -0.0003 & -0.008$ \pm $0.004 \\ 
\hline
2 & 0.0003 & -0.003$ \pm $0.003 \\ 
\hline
3 & -0.002 & 0.007$ \pm $0.006 \\ 
\hline
4 & -0.0027 & -0.013$ \pm $0.006 \\ 
\hline
5 & 0.9865 & 0.975$ \pm $0.006 \\ 
\hline
6 & 0.439 & 0.456$ \pm $0.009 \\ 
\hline
7 & 0.4398 & 0.439$ \pm $0.007 \\ 
\hline
8 & 0.9881 & 0.96$ \pm $0.004 \\ 
\hline
9 & -0.0033 & -0.008$ \pm $0.005 \\ 
\hline
10 & -0.0023 & 0.014$ \pm $0.005 \\ 
\hline
11 & 0.0016 & 0.01$ \pm $0.005 \\ 
\hline
\end{tabular}
\caption{The expectation of $\hat{P}_{0,x} \hat{P}_{1,x+1} + \hat{P}_{1,x} \hat{P}_{0,x+1}$ for $t=2.0$ displayed in Fig.\ref{fig:quarksIBM12}.}
\end{table}\begin{table}[]
\centering
\begin{tabular}{|c|c|c|}
\hline
$x$ & Classical Simulation & {\tt ibm\_torino} \\ 
\hline
0 & 0.0001 & 0.001$ \pm $0.004 \\ 
\hline
1 & -0.0001 & -0.0$ \pm $0.006 \\ 
\hline
2 & 0.0005 & -0.001$ \pm $0.005 \\ 
\hline
3 & 0.0009 & -0.0$ \pm $0.006 \\ 
\hline
4 & 0.0166 & -0.0$ \pm $0.008 \\ 
\hline
5 & 0.9984 & 0.986$ \pm $0.006 \\ 
\hline
6 & 0.6597 & 0.669$ \pm $0.012 \\ 
\hline
7 & 0.6598 & 0.663$ \pm $0.007 \\ 
\hline
8 & 0.9981 & 0.959$ \pm $0.008 \\ 
\hline
9 & 0.0169 & 0.0$ \pm $0.005 \\ 
\hline
10 & 0.0009 & 0.0$ \pm $0.004 \\ 
\hline
11 & 0.0 & 0.001$ \pm $0.005 \\ 
\hline
\end{tabular}
\caption{The expectation of $\hat{P}_{0,x} \hat{P}_{1,x+1} + \hat{P}_{1,x} \hat{P}_{0,x+1}$ for $t=3.0$ displayed in Fig.\ref{fig:quarksIBM12}.}
\end{table}\begin{table}[]
\centering
\begin{tabular}{|c|c|c|}
\hline
$x$ & Classical Simulation & {\tt ibm\_torino} \\ 
\hline
0 & -0.0026 & -0.001$ \pm $0.008 \\ 
\hline
1 & -0.0015 & -0.004$ \pm $0.01 \\ 
\hline
2 & -0.0005 & 0.0$ \pm $0.009 \\ 
\hline
3 & -0.0019 & -0.0$ \pm $0.01 \\ 
\hline
4 & 0.0226 & -0.0$ \pm $0.011 \\ 
\hline
5 & 0.9782 & 0.954$ \pm $0.009 \\ 
\hline
6 & 0.5672 & 0.596$ \pm $0.013 \\ 
\hline
7 & 0.5683 & 0.614$ \pm $0.01 \\ 
\hline
8 & 0.9782 & 0.948$ \pm $0.016 \\ 
\hline
9 & 0.019 & -0.022$ \pm $0.009 \\ 
\hline
10 & -0.0029 & 0.007$ \pm $0.012 \\ 
\hline
11 & -0.0001 & 0.002$ \pm $0.009 \\ 
\hline
\end{tabular}
\caption{The expectation of $\hat{P}_{0,x} \hat{P}_{1,x+1} + \hat{P}_{1,x} \hat{P}_{0,x+1}$ for $t=4.0$ displayed in Fig.\ref{fig:quarksIBM12}.}
\end{table}\begin{table}[]
\centering
\begin{tabular}{|c|c|c|}
\hline
$x$ & Classical Simulation & {\tt ibm\_torino} \\ 
\hline
0 & -0.0003 & -0.002$ \pm $0.011 \\ 
\hline
1 & 0.0019 & -0.0$ \pm $0.009 \\ 
\hline
2 & -0.0002 & -0.0$ \pm $0.009 \\ 
\hline
3 & 0.0007 & -0.0$ \pm $0.009 \\ 
\hline
4 & 0.046 & -0.0$ \pm $0.014 \\ 
\hline
5 & 0.9886 & 0.951$ \pm $0.008 \\ 
\hline
6 & 0.3149 & 0.383$ \pm $0.009 \\ 
\hline
7 & 0.3069 & 0.393$ \pm $0.01 \\ 
\hline
8 & 0.9866 & 0.936$ \pm $0.009 \\ 
\hline
9 & 0.0522 & -0.001$ \pm $0.01 \\ 
\hline
10 & 0.0026 & 0.001$ \pm $0.011 \\ 
\hline
11 & -0.0007 & -0.0$ \pm $0.011 \\ 
\hline
\end{tabular}
\caption{The expectation of $\hat{P}_{0,x} \hat{P}_{1,x+1} + \hat{P}_{1,x} \hat{P}_{0,x+1}$ for $t=5.0$ displayed in Fig.\ref{fig:quarksIBM12}.}
\end{table}\begin{table}[]
\centering
\begin{tabular}{|c|c|c|}
\hline
$x$ & Classical Simulation & {\tt ibm\_torino} \\ 
\hline
0 & 0.0004 & -0.0$ \pm $0.01 \\ 
\hline
1 & 0.0003 & 0.0$ \pm $0.014 \\ 
\hline
2 & 0.0003 & -0.003$ \pm $0.013 \\ 
\hline
3 & 0.0051 & 0.0$ \pm $0.016 \\ 
\hline
4 & 0.0641 & 0.001$ \pm $0.013 \\ 
\hline
5 & 0.9875 & 0.922$ \pm $0.009 \\ 
\hline
6 & 0.1018 & 0.216$ \pm $0.012 \\ 
\hline
7 & 0.0983 & 0.197$ \pm $0.015 \\ 
\hline
8 & 0.986 & 0.931$ \pm $0.019 \\ 
\hline
9 & 0.0677 & -0.0$ \pm $0.012 \\ 
\hline
10 & 0.007 & -0.0$ \pm $0.015 \\ 
\hline
11 & 0.0007 & -0.0$ \pm $0.008 \\ 
\hline
\end{tabular}
\caption{The expectation of $\hat{P}_{0,x} \hat{P}_{1,x+1} + \hat{P}_{1,x} \hat{P}_{0,x+1}$ for $t=6.0$ displayed in Fig.\ref{fig:quarksIBM12}.}
\end{table}\begin{table}[]
\centering
\begin{tabular}{|c|c|c|}
\hline
$x$ & Classical Simulation & {\tt ibm\_torino} \\ 
\hline
0 & 0.0029 & -0.001$ \pm $0.012 \\ 
\hline
1 & 0.0 & 0.0$ \pm $0.014 \\ 
\hline
2 & 0.0038 & -0.0$ \pm $0.018 \\ 
\hline
3 & 0.0107 & 0.001$ \pm $0.015 \\ 
\hline
4 & 0.0584 & 0.0$ \pm $0.018 \\ 
\hline
5 & 0.9611 & 0.947$ \pm $0.012 \\ 
\hline
6 & 0.0721 & 0.295$ \pm $0.015 \\ 
\hline
7 & 0.0718 & 0.315$ \pm $0.011 \\ 
\hline
8 & 0.9617 & 0.94$ \pm $0.018 \\ 
\hline
9 & 0.0607 & -0.0$ \pm $0.014 \\ 
\hline
10 & 0.0059 & 0.0$ \pm $0.019 \\ 
\hline
11 & 0.0005 & -0.0$ \pm $0.017 \\ 
\hline
\end{tabular}
\caption{The expectation of $\hat{P}_{0,x} \hat{P}_{1,x+1} + \hat{P}_{1,x} \hat{P}_{0,x+1}$ for $t=7.0$ displayed in Fig.\ref{fig:quarksIBM12}.}
\end{table}\begin{table}[]
\centering
\begin{tabular}{|c|c|c|}
\hline
$x$ & Classical Simulation & {\tt ibm\_torino} \\ 
\hline
0 & -0.0008 & 0.0$ \pm $0.017 \\ 
\hline
1 & -0.0003 & -0.0$ \pm $0.018 \\ 
\hline
2 & 0.0014 & 0.0$ \pm $0.024 \\ 
\hline
3 & 0.0148 & 0.001$ \pm $0.02 \\ 
\hline
4 & 0.0772 & -0.007$ \pm $0.015 \\ 
\hline
5 & 0.9674 & 0.938$ \pm $0.016 \\ 
\hline
6 & 0.2567 & 0.474$ \pm $0.019 \\ 
\hline
7 & 0.257 & 0.474$ \pm $0.02 \\ 
\hline
8 & 0.967 & 0.897$ \pm $0.016 \\ 
\hline
9 & 0.075 & 0.0$ \pm $0.021 \\ 
\hline
10 & 0.012 & 0.0$ \pm $0.017 \\ 
\hline
11 & -0.0004 & 0.0$ \pm $0.019 \\ 
\hline
\end{tabular}
\caption{The expectation of $\hat{P}_{0,x} \hat{P}_{1,x+1} + \hat{P}_{1,x} \hat{P}_{0,x+1}$ for $t=8.0$ displayed in Fig.\ref{fig:quarksIBM12}.}
\end{table}\begin{table}[]
\centering
\begin{tabular}{|c|c|c|}
\hline
$x$ & Classical Simulation & {\tt ibm\_torino} \\ 
\hline
0 & 0.0008 & 0.036$ \pm $0.162 \\ 
\hline
1 & 0.0004 & 0.005$ \pm $0.031 \\ 
\hline
2 & 0.0021 & 0.001$ \pm $0.146 \\ 
\hline
3 & 0.0189 & 0.0$ \pm $0.047 \\ 
\hline
4 & 0.0927 & -0.005$ \pm $0.04 \\ 
\hline
5 & 0.9569 & 0.806$ \pm $0.027 \\ 
\hline
6 & 0.4175 & 0.632$ \pm $0.047 \\ 
\hline
7 & 0.4173 & 0.578$ \pm $0.035 \\ 
\hline
8 & 0.9553 & 0.886$ \pm $0.055 \\ 
\hline
9 & 0.0934 & -0.0$ \pm $0.118 \\ 
\hline
10 & 0.0216 & 0.016$ \pm $0.139 \\ 
\hline
11 & 0.0031 & -0.079$ \pm $0.07 \\ 
\hline
\end{tabular}
\caption{The expectation of $\hat{P}_{0,x} \hat{P}_{1,x+1} + \hat{P}_{1,x} \hat{P}_{0,x+1}$ for $t=9.0$ displayed in Fig.\ref{fig:quarksIBM12}.}
\end{table}\begin{table}[]
\centering
\begin{tabular}{|c|c|c|}
\hline
$x$ & Classical Simulation & {\tt ibm\_torino} \\ 
\hline
0 & 0.0022 & 0.019$ \pm $0.24 \\ 
\hline
1 & 0.0047 & -0.027$ \pm $0.099 \\ 
\hline
2 & 0.0037 & -0.009$ \pm $0.156 \\ 
\hline
3 & 0.0226 & 0.06$ \pm $0.137 \\ 
\hline
4 & 0.1016 & 0.11$ \pm $0.024 \\ 
\hline
5 & 0.9353 & 0.754$ \pm $0.042 \\ 
\hline
6 & 0.4249 & 0.536$ \pm $0.113 \\ 
\hline
7 & 0.4242 & 0.422$ \pm $0.08 \\ 
\hline
8 & 0.9341 & 0.501$ \pm $0.287 \\ 
\hline
9 & 0.1 & 0.008$ \pm $0.091 \\ 
\hline
10 & 0.0215 & 0.129$ \pm $0.267 \\ 
\hline
11 & 0.0012 & 0.009$ \pm $0.37 \\ 
\hline
\end{tabular}
\caption{The expectation of $\hat{P}_{0,x} \hat{P}_{1,x+1} + \hat{P}_{1,x} \hat{P}_{0,x+1}$ for $t=10.0$ displayed in Fig.\ref{fig:quarksIBM12}.}
\end{table}\begin{table}[]
\centering
\begin{tabular}{|c|c|c|}
\hline
$x$ & Classical Simulation & {\tt ibm\_torino} \\ 
\hline
0 & -0.0008 & -0.161$ \pm $0.141 \\ 
\hline
1 & -0.0004 & 0.073$ \pm $0.119 \\ 
\hline
2 & 0.004 & 0.017$ \pm $0.129 \\ 
\hline
3 & 0.0375 & 0.0$ \pm $0.219 \\ 
\hline
4 & 0.1421 & 0.016$ \pm $0.101 \\ 
\hline
5 & 0.939 & 0.893$ \pm $0.178 \\ 
\hline
6 & 0.3249 & 0.494$ \pm $0.042 \\ 
\hline
7 & 0.3256 & 0.284$ \pm $0.286 \\ 
\hline
8 & 0.9379 & 0.584$ \pm $0.075 \\ 
\hline
9 & 0.1387 & -0.001$ \pm $0.041 \\ 
\hline
10 & 0.036 & 0.009$ \pm $0.365 \\ 
\hline
11 & 0.0037 & -0.097$ \pm $0.05 \\ 
\hline
\end{tabular}
\caption{The expectation of $\hat{P}_{0,x} \hat{P}_{1,x+1} + \hat{P}_{1,x} \hat{P}_{0,x+1}$ for $t=11.0$ displayed in Fig.\ref{fig:quarksIBM12}.}
\end{table}\begin{table}[]
\centering
\begin{tabular}{|c|c|c|}
\hline
$x$ & Classical Simulation & {\tt ibm\_torino} \\ 
\hline
0 & 0.0018 & -0.108$ \pm $0.034 \\ 
\hline
1 & 0.0018 & 0.005$ \pm $0.023 \\ 
\hline
2 & 0.0068 & 0.0$ \pm $0.038 \\ 
\hline
3 & 0.0444 & 0.001$ \pm $0.019 \\ 
\hline
4 & 0.1437 & 0.006$ \pm $0.025 \\ 
\hline
5 & 0.914 & 0.758$ \pm $0.018 \\ 
\hline
6 & 0.1838 & 0.499$ \pm $0.037 \\ 
\hline
7 & 0.1841 & 0.499$ \pm $0.022 \\ 
\hline
8 & 0.9198 & 0.792$ \pm $0.033 \\ 
\hline
9 & 0.1469 & 0.0$ \pm $0.028 \\ 
\hline
10 & 0.0426 & 0.004$ \pm $0.045 \\ 
\hline
11 & 0.0073 & -0.022$ \pm $0.035 \\ 
\hline
\end{tabular}
\caption{The expectation of $\hat{P}_{0,x} \hat{P}_{1,x+1} + \hat{P}_{1,x} \hat{P}_{0,x+1}$ for $t=12.0$ displayed in Fig.\ref{fig:quarksIBM12}.}
\end{table}\begin{table}[]
\centering
\begin{tabular}{|c|c|c|}
\hline
$x$ & Classical Simulation & {\tt ibm\_torino} \\ 
\hline
0 & -0.0007 & -0.021$ \pm $0.046 \\ 
\hline
1 & -0.0015 & -0.0$ \pm $0.04 \\ 
\hline
2 & 0.0094 & -0.002$ \pm $0.038 \\ 
\hline
3 & 0.0558 & 0.011$ \pm $0.033 \\ 
\hline
4 & 0.1479 & 0.004$ \pm $0.025 \\ 
\hline
5 & 0.8858 & 0.859$ \pm $0.032 \\ 
\hline
6 & 0.1364 & 0.47$ \pm $0.039 \\ 
\hline
7 & 0.1387 & 0.583$ \pm $0.028 \\ 
\hline
8 & 0.8898 & 0.817$ \pm $0.036 \\ 
\hline
9 & 0.1488 & 0.002$ \pm $0.053 \\ 
\hline
10 & 0.0564 & -0.0$ \pm $0.026 \\ 
\hline
11 & 0.0116 & 0.062$ \pm $0.032 \\ 
\hline
\end{tabular}
\caption{The expectation of $\hat{P}_{0,x} \hat{P}_{1,x+1} + \hat{P}_{1,x} \hat{P}_{0,x+1}$ for $t=13.0$ displayed in Fig.\ref{fig:quarksIBM12}.}
\end{table}

\FloatBarrier

\begin{longtable}{|c|c|c|}
\hline
$x$ & Classical Simulation & {\tt ibm\_torino} \\ 
\hline
0 & 0.0007 & 0.048$ \pm $0.007 \\ 
\hline
1 & 0.0016 & -0.007$ \pm $0.01 \\ 
\hline
2 & 0.0006 & -0.017$ \pm $0.006 \\ 
\hline
3 & -0.001 & -0.009$ \pm $0.008 \\ 
\hline
4 & 0.0013 & 0.011$ \pm $0.008 \\ 
\hline
5 & 0.0029 & 0.009$ \pm $0.003 \\ 
\hline
6 & 0.0004 & 0.001$ \pm $0.01 \\ 
\hline
7 & -0.0007 & -0.062$ \pm $0.04 \\ 
\hline
8 & 0.0001 & -0.054$ \pm $0.004 \\ 
\hline
9 & -0.0002 & 0.065$ \pm $0.061 \\ 
\hline
10 & -0.001 & 0.094$ \pm $0.008 \\ 
\hline
11 & -0.0029 & 0.005$ \pm $0.005 \\ 
\hline
12 & -0.0033 & 0.006$ \pm $0.006 \\ 
\hline
13 & -0.0007 & 0.012$ \pm $0.005 \\ 
\hline
14 & -0.0028 & -0.0$ \pm $0.004 \\ 
\hline
15 & -0.0036 & 0.012$ \pm $0.006 \\ 
\hline
16 & -0.0002 & 0.001$ \pm $0.008 \\ 
\hline
17 & 0.001 & 0.001$ \pm $0.007 \\ 
\hline
18 & 0.0014 & 0.001$ \pm $0.006 \\ 
\hline
19 & 0.0018 & 0.006$ \pm $0.004 \\ 
\hline
20 & 0.0004 & -0.01$ \pm $0.006 \\ 
\hline
21 & -0.0007 & -0.012$ \pm $0.005 \\ 
\hline
22 & -0.0012 & 0.008$ \pm $0.004 \\ 
\hline
23 & -0.003 & -0.006$ \pm $0.007 \\ 
\hline
24 & -0.0016 & -0.014$ \pm $0.007 \\ 
\hline
25 & 0.0005 & -0.055$ \pm $0.058 \\ 
\hline
26 & 0.0008 & -0.049$ \pm $0.006 \\ 
\hline
27 & -0.003 & -0.062$ \pm $0.052 \\ 
\hline
28 & -0.0106 & -0.072$ \pm $0.007 \\ 
\hline
29 & 0.9832 & 0.97$ \pm $0.01 \\ 
\hline
30 & 0.1198 & 0.112$ \pm $0.015 \\ 
\hline
31 & 0.2491 & 0.167$ \pm $0.06 \\ 
\hline
32 & 0.2366 & -0.013$ \pm $0.053 \\ 
\hline
33 & 0.2324 & 0.036$ \pm $0.066 \\ 
\hline
34 & 0.232 & 0.131$ \pm $0.013 \\ 
\hline
35 & 0.2273 & 0.186$ \pm $0.009 \\ 
\hline
36 & 0.23 & 0.174$ \pm $0.04 \\ 
\hline
37 & 0.2332 & 0.168$ \pm $0.011 \\ 
\hline
38 & 0.2331 & 0.256$ \pm $0.056 \\ 
\hline
39 & 0.236 & 0.252$ \pm $0.012 \\ 
\hline
40 & 0.2358 & 0.213$ \pm $0.015 \\ 
\hline
41 & 0.2309 & 0.233$ \pm $0.032 \\ 
\hline
42 & 0.2277 & 0.237$ \pm $0.048 \\ 
\hline
43 & 0.226 & 0.326$ \pm $0.072 \\ 
\hline
44 & 0.2356 & 0.276$ \pm $0.009 \\ 
\hline
45 & 0.2388 & 0.194$ \pm $0.041 \\ 
\hline
46 & 0.2349 & 0.191$ \pm $0.013 \\ 
\hline
47 & 0.2369 & 0.247$ \pm $0.065 \\ 
\hline
48 & 0.2351 & 0.222$ \pm $0.011 \\ 
\hline
49 & 0.2374 & 0.223$ \pm $0.07 \\ 
\hline
50 & 0.235 & 0.213$ \pm $0.036 \\ 
\hline
51 & 0.2357 & 0.232$ \pm $0.043 \\ 
\hline
52 & 0.2363 & 0.235$ \pm $0.034 \\ 
\hline
53 & 0.2379 & 0.16$ \pm $0.013 \\ 
\hline
54 & 0.2444 & 0.208$ \pm $0.041 \\ 
\hline
55 & 0.2349 & 0.206$ \pm $0.01 \\ 
\hline
56 & 0.2295 & 0.196$ \pm $0.058 \\ 
\hline
57 & 0.2285 & 0.19$ \pm $0.011 \\ 
\hline
58 & 0.2298 & 0.209$ \pm $0.009 \\ 
\hline
59 & 0.2414 & 0.235$ \pm $0.012 \\ 
\hline
60 & 0.242 & 0.285$ \pm $0.045 \\ 
\hline
61 & 0.2342 & 0.295$ \pm $0.01 \\ 
\hline
62 & 0.2342 & 0.212$ \pm $0.009 \\ 
\hline
63 & 0.234 & 0.193$ \pm $0.018 \\ 
\hline
64 & 0.2334 & 0.241$ \pm $0.014 \\ 
\hline
65 & 0.2347 & 0.244$ \pm $0.009 \\ 
\hline
66 & 0.2404 & 0.259$ \pm $0.011 \\ 
\hline
67 & 0.2367 & 0.265$ \pm $0.009 \\ 
\hline
68 & 0.2415 & 0.189$ \pm $0.035 \\ 
\hline
69 & 0.1207 & 0.064$ \pm $0.005 \\ 
\hline
70 & 0.9837 & 0.994$ \pm $0.006 \\ 
\hline
71 & -0.0074 & -0.004$ \pm $0.004 \\ 
\hline
72 & -0.0036 & -0.011$ \pm $0.007 \\ 
\hline
73 & -0.0009 & 0.006$ \pm $0.034 \\ 
\hline
74 & -0.0001 & -0.0$ \pm $0.046 \\ 
\hline
75 & -0.0015 & -0.036$ \pm $0.004 \\ 
\hline
76 & -0.0001 & 0.004$ \pm $0.004 \\ 
\hline
77 & -0.0004 & 0.003$ \pm $0.009 \\ 
\hline
78 & 0.0008 & 0.003$ \pm $0.004 \\ 
\hline
79 & 0.0006 & -0.028$ \pm $0.005 \\ 
\hline
80 & 0.0012 & 0.002$ \pm $0.01 \\ 
\hline
81 & 0.0029 & -0.001$ \pm $0.005 \\ 
\hline
82 & -0.0001 & -0.002$ \pm $0.007 \\ 
\hline
83 & -0.0007 & -0.054$ \pm $0.047 \\ 
\hline
84 & 0.0017 & -0.064$ \pm $0.005 \\ 
\hline
85 & 0.001 & 0.042$ \pm $0.054 \\ 
\hline
86 & -0.0002 & 0.065$ \pm $0.008 \\ 
\hline
87 & -0.0005 & 0.025$ \pm $0.05 \\ 
\hline
88 & -0.0007 & 0.019$ \pm $0.006 \\ 
\hline
89 & -0.0014 & -0.057$ \pm $0.05 \\ 
\hline
90 & -0.0002 & -0.061$ \pm $0.006 \\ 
\hline
91 & 0.0008 & 0.009$ \pm $0.011 \\ 
\hline
92 & -0.0001 & 0.037$ \pm $0.05 \\ 
\hline
93 & -0.0022 & 0.133$ \pm $0.068 \\ 
\hline
94 & -0.0019 & 0.094$ \pm $0.004 \\ 
\hline
95 & 0.0003 & -0.0$ \pm $0.006 \\ 
\hline
96 & 0.0 & 0.003$ \pm $0.06 \\ 
\hline
97 & 0.0019 & -0.003$ \pm $0.012 \\ 
\hline
98 & 0.0025 & 0.054$ \pm $0.047 \\ 
\hline
99 & -0.0022 & 0.058$ \pm $0.008 \\ 
\hline
100 & -0.003 & -0.013$ \pm $0.061 \\ 
\hline
101 & 0.0 & -0.009$ \pm $0.006 \\ 
\hline
102 & -0.0019 & -0.021$ \pm $0.04 \\ 
\hline
103 & -0.0021 & 0.002$ \pm $0.05 \\ 
\hline
\caption{The expectation of $\hat{P}_{0,x} \hat{P}_{1,x+1} + \hat{P}_{1,x} \hat{P}_{0,x+1}$ for $t=1.0$ displayed in Fig.\ref{fig:quarksIBM104}.}
\end{longtable}
\begin{longtable}{|c|c|c|}
\hline
$x$ & Classical Simulation & {\tt ibm\_torino} \\ 
\hline
0 & -0.0013 & -0.003$ \pm $0.007 \\ 
\hline
1 & 0.0002 & 0.022$ \pm $0.016 \\ 
\hline
2 & 0.0004 & 0.018$ \pm $0.01 \\ 
\hline
3 & -0.0001 & -0.016$ \pm $0.01 \\ 
\hline
4 & 0.0001 & -0.035$ \pm $0.031 \\ 
\hline
5 & 0.0003 & -0.055$ \pm $0.051 \\ 
\hline
6 & 0.0029 & 0.002$ \pm $0.01 \\ 
\hline
7 & 0.0026 & 0.018$ \pm $0.014 \\ 
\hline
8 & 0.0011 & -0.044$ \pm $0.013 \\ 
\hline
9 & -0.0008 & 0.01$ \pm $0.015 \\ 
\hline
10 & -0.0015 & -0.015$ \pm $0.024 \\ 
\hline
11 & -0.0013 & -0.005$ \pm $0.015 \\ 
\hline
12 & -0.0026 & 0.05$ \pm $0.066 \\ 
\hline
13 & -0.0011 & 0.047$ \pm $0.017 \\ 
\hline
14 & 0.0013 & -0.015$ \pm $0.061 \\ 
\hline
15 & -0.0003 & -0.004$ \pm $0.013 \\ 
\hline
16 & -0.0005 & 0.015$ \pm $0.01 \\ 
\hline
17 & 0.0009 & 0.004$ \pm $0.009 \\ 
\hline
18 & 0.0017 & -0.024$ \pm $0.009 \\ 
\hline
19 & 0.0015 & -0.021$ \pm $0.006 \\ 
\hline
20 & 0.0009 & -0.04$ \pm $0.013 \\ 
\hline
21 & 0.0011 & -0.053$ \pm $0.052 \\ 
\hline
22 & 0.0005 & -0.018$ \pm $0.009 \\ 
\hline
23 & 0.0004 & 0.008$ \pm $0.028 \\ 
\hline
24 & -0.0003 & -0.011$ \pm $0.051 \\ 
\hline
25 & 0.001 & -0.028$ \pm $0.017 \\ 
\hline
26 & 0.0007 & 0.072$ \pm $0.04 \\ 
\hline
27 & -0.0011 & 0.062$ \pm $0.01 \\ 
\hline
28 & -0.0026 & -0.001$ \pm $0.009 \\ 
\hline
29 & 0.9857 & 0.98$ \pm $0.008 \\ 
\hline
30 & 0.3587 & 0.377$ \pm $0.011 \\ 
\hline
31 & 0.6821 & 0.662$ \pm $0.02 \\ 
\hline
32 & 0.6297 & 0.653$ \pm $0.061 \\ 
\hline
33 & 0.634 & 0.687$ \pm $0.062 \\ 
\hline
34 & 0.644 & 0.616$ \pm $0.074 \\ 
\hline
35 & 0.6415 & 0.525$ \pm $0.048 \\ 
\hline
36 & 0.6405 & 0.53$ \pm $0.02 \\ 
\hline
37 & 0.6445 & 0.633$ \pm $0.024 \\ 
\hline
38 & 0.6347 & 0.668$ \pm $0.014 \\ 
\hline
39 & 0.6418 & 0.629$ \pm $0.015 \\ 
\hline
40 & 0.6421 & 0.649$ \pm $0.016 \\ 
\hline
41 & 0.6402 & 0.682$ \pm $0.014 \\ 
\hline
42 & 0.6396 & 0.648$ \pm $0.008 \\ 
\hline
43 & 0.6423 & 0.573$ \pm $0.022 \\ 
\hline
44 & 0.6467 & 0.504$ \pm $0.083 \\ 
\hline
45 & 0.6449 & 0.484$ \pm $0.025 \\ 
\hline
46 & 0.6388 & 0.754$ \pm $0.049 \\ 
\hline
47 & 0.6336 & 0.818$ \pm $0.022 \\ 
\hline
48 & 0.6401 & 0.638$ \pm $0.022 \\ 
\hline
49 & 0.6374 & 0.614$ \pm $0.039 \\ 
\hline
50 & 0.64 & 0.663$ \pm $0.022 \\ 
\hline
51 & 0.6445 & 0.593$ \pm $0.015 \\ 
\hline
52 & 0.638 & 0.66$ \pm $0.045 \\ 
\hline
53 & 0.6432 & 0.638$ \pm $0.024 \\ 
\hline
54 & 0.6466 & 0.63$ \pm $0.069 \\ 
\hline
55 & 0.643 & 0.629$ \pm $0.022 \\ 
\hline
56 & 0.6431 & 0.638$ \pm $0.042 \\ 
\hline
57 & 0.6425 & 0.638$ \pm $0.013 \\ 
\hline
58 & 0.6431 & 0.666$ \pm $0.022 \\ 
\hline
59 & 0.6369 & 0.702$ \pm $0.024 \\ 
\hline
60 & 0.6423 & 0.679$ \pm $0.022 \\ 
\hline
61 & 0.6414 & 0.654$ \pm $0.019 \\ 
\hline
62 & 0.6406 & 0.624$ \pm $0.014 \\ 
\hline
63 & 0.6444 & 0.631$ \pm $0.011 \\ 
\hline
64 & 0.6456 & 0.611$ \pm $0.015 \\ 
\hline
65 & 0.648 & 0.633$ \pm $0.041 \\ 
\hline
66 & 0.6413 & 0.692$ \pm $0.057 \\ 
\hline
67 & 0.6327 & 0.71$ \pm $0.073 \\ 
\hline
68 & 0.6828 & 0.801$ \pm $0.063 \\ 
\hline
69 & 0.3604 & 0.282$ \pm $0.03 \\ 
\hline
70 & 0.9874 & 0.902$ \pm $0.024 \\ 
\hline
71 & -0.0003 & 0.026$ \pm $0.009 \\ 
\hline
72 & 0.0007 & 0.005$ \pm $0.01 \\ 
\hline
73 & -0.0019 & 0.003$ \pm $0.01 \\ 
\hline
74 & -0.0027 & -0.009$ \pm $0.008 \\ 
\hline
75 & -0.0012 & -0.02$ \pm $0.01 \\ 
\hline
76 & -0.001 & -0.021$ \pm $0.01 \\ 
\hline
77 & -0.0011 & -0.005$ \pm $0.072 \\ 
\hline
78 & -0.0003 & 0.014$ \pm $0.011 \\ 
\hline
79 & 0.0016 & -0.053$ \pm $0.056 \\ 
\hline
80 & 0.002 & -0.079$ \pm $0.011 \\ 
\hline
81 & 0.0005 & -0.016$ \pm $0.006 \\ 
\hline
82 & 0.0009 & 0.006$ \pm $0.007 \\ 
\hline
83 & 0.0019 & -0.005$ \pm $0.012 \\ 
\hline
84 & 0.0001 & 0.005$ \pm $0.01 \\ 
\hline
85 & -0.0002 & -0.017$ \pm $0.068 \\ 
\hline
86 & -0.0001 & -0.018$ \pm $0.054 \\ 
\hline
87 & -0.0018 & 0.098$ \pm $0.025 \\ 
\hline
88 & 0.0016 & 0.042$ \pm $0.076 \\ 
\hline
89 & 0.0013 & 0.04$ \pm $0.013 \\ 
\hline
90 & -0.0034 & -0.008$ \pm $0.005 \\ 
\hline
91 & -0.0025 & -0.014$ \pm $0.013 \\ 
\hline
92 & -0.0004 & -0.008$ \pm $0.011 \\ 
\hline
93 & -0.0016 & -0.004$ \pm $0.011 \\ 
\hline
94 & -0.0026 & 0.068$ \pm $0.1 \\ 
\hline
95 & -0.002 & 0.077$ \pm $0.012 \\ 
\hline
96 & -0.0004 & 0.066$ \pm $0.043 \\ 
\hline
97 & -0.0005 & 0.03$ \pm $0.019 \\ 
\hline
98 & -0.0004 & -0.034$ \pm $0.008 \\ 
\hline
99 & -0.0001 & -0.005$ \pm $0.013 \\ 
\hline
100 & 0.0002 & -0.013$ \pm $0.016 \\ 
\hline
101 & -0.0011 & 0.009$ \pm $0.016 \\ 
\hline
102 & -0.0024 & 0.024$ \pm $0.011 \\ 
\hline
103 & -0.001 & -0.001$ \pm $0.009 \\ 
\hline
\caption{The expectation of $\hat{P}_{0,x} \hat{P}_{1,x+1} + \hat{P}_{1,x} \hat{P}_{0,x+1}$ for $t=2.0$ displayed in Fig.\ref{fig:quarksIBM104}.}
\end{longtable}
\begin{longtable}{|c|c|c|}
\hline
$x$ & Classical Simulation & {\tt ibm\_torino} \\ 
\hline
0 & 0.0013 & 0.0$ \pm $0.025 \\ 
\hline
1 & 0.0004 & 0.001$ \pm $0.017 \\ 
\hline
2 & 0.0002 & 0.059$ \pm $0.064 \\ 
\hline
3 & 0.0005 & 0.057$ \pm $0.021 \\ 
\hline
4 & 0.0003 & 0.001$ \pm $0.01 \\ 
\hline
5 & -0.0001 & -0.013$ \pm $0.072 \\ 
\hline
6 & -0.0002 & -0.04$ \pm $0.051 \\ 
\hline
7 & 0.0 & 0.03$ \pm $0.034 \\ 
\hline
8 & 0.0003 & 0.001$ \pm $0.057 \\ 
\hline
9 & 0.0 & -0.022$ \pm $0.049 \\ 
\hline
10 & 0.0001 & -0.01$ \pm $0.016 \\ 
\hline
11 & 0.0004 & -0.018$ \pm $0.021 \\ 
\hline
12 & 0.0005 & 0.007$ \pm $0.009 \\ 
\hline
13 & 0.0006 & 0.002$ \pm $0.017 \\ 
\hline
14 & 0.0003 & 0.024$ \pm $0.051 \\ 
\hline
15 & 0.0004 & 0.004$ \pm $0.029 \\ 
\hline
16 & 0.0001 & -0.003$ \pm $0.042 \\ 
\hline
17 & 0.0 & 0.028$ \pm $0.009 \\ 
\hline
18 & 0.0 & -0.0$ \pm $0.013 \\ 
\hline
19 & -0.0003 & 0.001$ \pm $0.01 \\ 
\hline
20 & 0.0002 & -0.008$ \pm $0.012 \\ 
\hline
21 & -0.0001 & 0.018$ \pm $0.049 \\ 
\hline
22 & -0.0001 & 0.037$ \pm $0.017 \\ 
\hline
23 & -0.0003 & -0.021$ \pm $0.05 \\ 
\hline
24 & -0.0003 & -0.028$ \pm $0.014 \\ 
\hline
25 & 0.0006 & 0.002$ \pm $0.01 \\ 
\hline
26 & 0.0008 & 0.0$ \pm $0.003 \\ 
\hline
27 & 0.0007 & -0.015$ \pm $0.042 \\ 
\hline
28 & 0.0148 & -0.032$ \pm $0.011 \\ 
\hline
29 & 0.9991 & 0.923$ \pm $0.011 \\ 
\hline
30 & 0.4492 & 0.485$ \pm $0.019 \\ 
\hline
31 & 0.8023 & 0.878$ \pm $0.024 \\ 
\hline
32 & 0.6948 & 0.788$ \pm $0.014 \\ 
\hline
33 & 0.7111 & 0.798$ \pm $0.017 \\ 
\hline
34 & 0.71 & 0.735$ \pm $0.05 \\ 
\hline
35 & 0.7116 & 0.807$ \pm $0.033 \\ 
\hline
36 & 0.7058 & 0.778$ \pm $0.085 \\ 
\hline
37 & 0.7096 & 0.696$ \pm $0.018 \\ 
\hline
38 & 0.7104 & 0.723$ \pm $0.021 \\ 
\hline
39 & 0.7066 & 0.747$ \pm $0.019 \\ 
\hline
40 & 0.7142 & 0.753$ \pm $0.015 \\ 
\hline
41 & 0.7076 & 0.735$ \pm $0.019 \\ 
\hline
42 & 0.7122 & 0.762$ \pm $0.052 \\ 
\hline
43 & 0.7155 & 0.674$ \pm $0.025 \\ 
\hline
44 & 0.7087 & 0.759$ \pm $0.06 \\ 
\hline
45 & 0.7075 & 0.791$ \pm $0.022 \\ 
\hline
46 & 0.7136 & 0.702$ \pm $0.023 \\ 
\hline
47 & 0.719 & 0.741$ \pm $0.028 \\ 
\hline
48 & 0.7136 & 0.667$ \pm $0.085 \\ 
\hline
49 & 0.7116 & 0.641$ \pm $0.06 \\ 
\hline
50 & 0.705 & 0.608$ \pm $0.026 \\ 
\hline
51 & 0.7089 & 0.65$ \pm $0.048 \\ 
\hline
52 & 0.7096 & 0.674$ \pm $0.026 \\ 
\hline
53 & 0.7009 & 0.703$ \pm $0.069 \\ 
\hline
54 & 0.7056 & 0.724$ \pm $0.03 \\ 
\hline
55 & 0.6979 & 0.71$ \pm $0.044 \\ 
\hline
56 & 0.6934 & 0.608$ \pm $0.016 \\ 
\hline
57 & 0.7038 & 0.731$ \pm $0.02 \\ 
\hline
58 & 0.7046 & 0.788$ \pm $0.037 \\ 
\hline
59 & 0.7033 & 0.719$ \pm $0.02 \\ 
\hline
60 & 0.711 & 0.734$ \pm $0.016 \\ 
\hline
61 & 0.7081 & 0.729$ \pm $0.016 \\ 
\hline
62 & 0.7039 & 0.712$ \pm $0.048 \\ 
\hline
63 & 0.6975 & 0.756$ \pm $0.021 \\ 
\hline
64 & 0.6994 & 0.694$ \pm $0.023 \\ 
\hline
65 & 0.7156 & 0.756$ \pm $0.078 \\ 
\hline
66 & 0.714 & 0.736$ \pm $0.034 \\ 
\hline
67 & 0.6869 & 0.801$ \pm $0.05 \\ 
\hline
68 & 0.7941 & 0.931$ \pm $0.017 \\ 
\hline
69 & 0.4501 & 0.523$ \pm $0.026 \\ 
\hline
70 & 0.9984 & 0.92$ \pm $0.01 \\ 
\hline
71 & 0.0128 & 0.0$ \pm $0.011 \\ 
\hline
72 & 0.0013 & -0.009$ \pm $0.017 \\ 
\hline
73 & 0.0003 & -0.046$ \pm $0.013 \\ 
\hline
74 & 0.0 & 0.02$ \pm $0.034 \\ 
\hline
75 & 0.0004 & 0.027$ \pm $0.016 \\ 
\hline
76 & 0.0007 & 0.049$ \pm $0.058 \\ 
\hline
77 & -0.0004 & 0.078$ \pm $0.017 \\ 
\hline
78 & -0.0006 & -0.002$ \pm $0.025 \\ 
\hline
79 & -0.0005 & -0.005$ \pm $0.011 \\ 
\hline
80 & -0.0001 & 0.005$ \pm $0.028 \\ 
\hline
81 & 0.0005 & 0.028$ \pm $0.011 \\ 
\hline
82 & 0.0001 & 0.013$ \pm $0.018 \\ 
\hline
83 & -0.0001 & -0.001$ \pm $0.016 \\ 
\hline
84 & 0.0002 & -0.014$ \pm $0.015 \\ 
\hline
85 & 0.0002 & 0.023$ \pm $0.06 \\ 
\hline
86 & 0.0001 & 0.044$ \pm $0.008 \\ 
\hline
87 & 0.0001 & -0.0$ \pm $0.013 \\ 
\hline
88 & -0.0003 & -0.012$ \pm $0.008 \\ 
\hline
89 & -0.0004 & -0.001$ \pm $0.017 \\ 
\hline
90 & -0.0001 & -0.002$ \pm $0.017 \\ 
\hline
91 & -0.0002 & 0.011$ \pm $0.024 \\ 
\hline
92 & -0.0005 & -0.012$ \pm $0.018 \\ 
\hline
93 & -0.0006 & -0.017$ \pm $0.01 \\ 
\hline
94 & -0.0003 & 0.007$ \pm $0.021 \\ 
\hline
95 & 0.0001 & 0.018$ \pm $0.023 \\ 
\hline
96 & 0.0 & -0.008$ \pm $0.044 \\ 
\hline
97 & -0.0002 & -0.102$ \pm $0.085 \\ 
\hline
98 & 0.0001 & -0.066$ \pm $0.019 \\ 
\hline
99 & -0.0004 & -0.026$ \pm $0.026 \\ 
\hline
100 & -0.0005 & -0.024$ \pm $0.009 \\ 
\hline
101 & -0.0001 & -0.004$ \pm $0.01 \\ 
\hline
102 & -0.0002 & -0.015$ \pm $0.009 \\ 
\hline
103 & 0.0009 & -0.011$ \pm $0.02 \\ 
\hline
\caption{The expectation of $\hat{P}_{0,x} \hat{P}_{1,x+1} + \hat{P}_{1,x} \hat{P}_{0,x+1}$ for $t=3.0$ displayed in Fig.\ref{fig:quarksIBM104}.}
\end{longtable}
\begin{longtable}{|c|c|c|}
\hline
$x$ & Classical Simulation & {\tt ibm\_torino} \\ 
\hline
0 & -0.0004 & 0.001$ \pm $0.014 \\ 
\hline
1 & -0.0011 & -0.003$ \pm $0.013 \\ 
\hline
2 & 0.0016 & -0.009$ \pm $0.013 \\ 
\hline
3 & -0.0005 & 0.003$ \pm $0.018 \\ 
\hline
4 & -0.0033 & -0.011$ \pm $0.02 \\ 
\hline
5 & -0.0021 & -0.014$ \pm $0.019 \\ 
\hline
6 & 0.001 & -0.002$ \pm $0.012 \\ 
\hline
7 & 0.0 & -0.002$ \pm $0.014 \\ 
\hline
8 & -0.0009 & 0.001$ \pm $0.008 \\ 
\hline
9 & -0.0022 & 0.0$ \pm $0.017 \\ 
\hline
10 & -0.0036 & -0.033$ \pm $0.015 \\ 
\hline
11 & 0.0015 & -0.023$ \pm $0.011 \\ 
\hline
12 & 0.0014 & 0.01$ \pm $0.012 \\ 
\hline
13 & -0.0005 & 0.001$ \pm $0.011 \\ 
\hline
14 & 0.0009 & -0.005$ \pm $0.012 \\ 
\hline
15 & 0.0015 & -0.005$ \pm $0.014 \\ 
\hline
16 & 0.0 & -0.003$ \pm $0.012 \\ 
\hline
17 & 0.0 & -0.007$ \pm $0.016 \\ 
\hline
18 & 0.0008 & 0.009$ \pm $0.013 \\ 
\hline
19 & -0.0019 & -0.009$ \pm $0.017 \\ 
\hline
20 & -0.005 & -0.015$ \pm $0.015 \\ 
\hline
21 & -0.0035 & -0.012$ \pm $0.012 \\ 
\hline
22 & 0.0004 & 0.009$ \pm $0.011 \\ 
\hline
23 & -0.0015 & -0.005$ \pm $0.016 \\ 
\hline
24 & -0.0017 & 0.007$ \pm $0.016 \\ 
\hline
25 & -0.0004 & 0.01$ \pm $0.02 \\ 
\hline
26 & -0.0037 & -0.001$ \pm $0.009 \\ 
\hline
27 & -0.0042 & -0.017$ \pm $0.014 \\ 
\hline
28 & 0.014 & -0.003$ \pm $0.014 \\ 
\hline
29 & 0.9788 & 0.973$ \pm $0.013 \\ 
\hline
30 & 0.2886 & 0.462$ \pm $0.016 \\ 
\hline
31 & 0.4824 & 0.609$ \pm $0.014 \\ 
\hline
32 & 0.3511 & 0.516$ \pm $0.012 \\ 
\hline
33 & 0.3848 & 0.528$ \pm $0.014 \\ 
\hline
34 & 0.3829 & 0.534$ \pm $0.014 \\ 
\hline
35 & 0.3852 & 0.496$ \pm $0.016 \\ 
\hline
36 & 0.3849 & 0.504$ \pm $0.017 \\ 
\hline
37 & 0.3832 & 0.534$ \pm $0.014 \\ 
\hline
38 & 0.3836 & 0.534$ \pm $0.011 \\ 
\hline
39 & 0.3842 & 0.552$ \pm $0.017 \\ 
\hline
40 & 0.3771 & 0.493$ \pm $0.018 \\ 
\hline
41 & 0.3793 & 0.495$ \pm $0.013 \\ 
\hline
42 & 0.3896 & 0.53$ \pm $0.015 \\ 
\hline
43 & 0.3835 & 0.495$ \pm $0.018 \\ 
\hline
44 & 0.3817 & 0.551$ \pm $0.018 \\ 
\hline
45 & 0.3807 & 0.535$ \pm $0.013 \\ 
\hline
46 & 0.3756 & 0.553$ \pm $0.016 \\ 
\hline
47 & 0.3817 & 0.552$ \pm $0.014 \\ 
\hline
48 & 0.377 & 0.543$ \pm $0.017 \\ 
\hline
49 & 0.3744 & 0.537$ \pm $0.017 \\ 
\hline
50 & 0.3808 & 0.534$ \pm $0.012 \\ 
\hline
51 & 0.3841 & 0.537$ \pm $0.013 \\ 
\hline
52 & 0.3888 & 0.544$ \pm $0.012 \\ 
\hline
53 & 0.38 & 0.506$ \pm $0.012 \\ 
\hline
54 & 0.3777 & 0.497$ \pm $0.018 \\ 
\hline
55 & 0.3816 & 0.546$ \pm $0.016 \\ 
\hline
56 & 0.3789 & 0.507$ \pm $0.016 \\ 
\hline
57 & 0.3803 & 0.507$ \pm $0.014 \\ 
\hline
58 & 0.3807 & 0.537$ \pm $0.014 \\ 
\hline
59 & 0.3763 & 0.524$ \pm $0.011 \\ 
\hline
60 & 0.3808 & 0.544$ \pm $0.014 \\ 
\hline
61 & 0.3844 & 0.509$ \pm $0.014 \\ 
\hline
62 & 0.3846 & 0.542$ \pm $0.015 \\ 
\hline
63 & 0.3844 & 0.514$ \pm $0.016 \\ 
\hline
64 & 0.3841 & 0.543$ \pm $0.013 \\ 
\hline
65 & 0.3813 & 0.547$ \pm $0.016 \\ 
\hline
66 & 0.3827 & 0.542$ \pm $0.014 \\ 
\hline
67 & 0.3488 & 0.513$ \pm $0.015 \\ 
\hline
68 & 0.4799 & 0.605$ \pm $0.019 \\ 
\hline
69 & 0.2884 & 0.426$ \pm $0.014 \\ 
\hline
70 & 0.9803 & 0.95$ \pm $0.012 \\ 
\hline
71 & 0.0151 & -0.007$ \pm $0.011 \\ 
\hline
72 & -0.0041 & -0.008$ \pm $0.011 \\ 
\hline
73 & 0.0004 & -0.002$ \pm $0.015 \\ 
\hline
74 & 0.0004 & -0.007$ \pm $0.012 \\ 
\hline
75 & -0.0003 & -0.005$ \pm $0.014 \\ 
\hline
76 & -0.0003 & 0.0$ \pm $0.012 \\ 
\hline
77 & -0.0026 & -0.001$ \pm $0.012 \\ 
\hline
78 & -0.0017 & -0.001$ \pm $0.012 \\ 
\hline
79 & 0.0001 & 0.007$ \pm $0.013 \\ 
\hline
80 & 0.0001 & -0.001$ \pm $0.011 \\ 
\hline
81 & 0.0004 & -0.0$ \pm $0.013 \\ 
\hline
82 & 0.001 & -0.005$ \pm $0.015 \\ 
\hline
83 & 0.0024 & 0.01$ \pm $0.023 \\ 
\hline
84 & 0.0028 & -0.038$ \pm $0.018 \\ 
\hline
85 & 0.001 & 0.006$ \pm $0.013 \\ 
\hline
86 & -0.0036 & -0.003$ \pm $0.015 \\ 
\hline
87 & -0.0033 & -0.009$ \pm $0.017 \\ 
\hline
88 & -0.0006 & 0.0$ \pm $0.014 \\ 
\hline
89 & -0.0016 & -0.006$ \pm $0.019 \\ 
\hline
90 & -0.0008 & -0.002$ \pm $0.013 \\ 
\hline
91 & -0.0003 & -0.001$ \pm $0.016 \\ 
\hline
92 & 0.0002 & 0.001$ \pm $0.016 \\ 
\hline
93 & 0.0019 & 0.001$ \pm $0.016 \\ 
\hline
94 & 0.0005 & 0.004$ \pm $0.013 \\ 
\hline
95 & 0.0009 & 0.018$ \pm $0.013 \\ 
\hline
96 & -0.0001 & 0.003$ \pm $0.012 \\ 
\hline
97 & -0.0026 & -0.002$ \pm $0.012 \\ 
\hline
98 & 0.0001 & -0.004$ \pm $0.014 \\ 
\hline
99 & 0.0026 & -0.0$ \pm $0.014 \\ 
\hline
100 & 0.0036 & -0.001$ \pm $0.014 \\ 
\hline
101 & 0.0041 & -0.007$ \pm $0.015 \\ 
\hline
102 & 0.0027 & -0.001$ \pm $0.016 \\ 
\hline
103 & 0.002 & 0.011$ \pm $0.016 \\ 
\hline
\caption{The expectation of $\hat{P}_{0,x} \hat{P}_{1,x+1} + \hat{P}_{1,x} \hat{P}_{0,x+1}$ for $t=4.0$ displayed in Fig.\ref{fig:quarksIBM104}.}
\end{longtable}
\begin{longtable}{|c|c|c|}
\hline
$x$ & Classical Simulation & {\tt ibm\_torino} \\ 
\hline
0 & -0.0007 & -0.007$ \pm $0.015 \\ 
\hline
1 & 0.0004 & 0.002$ \pm $0.02 \\ 
\hline
2 & -0.001 & 0.0$ \pm $0.011 \\ 
\hline
3 & -0.0001 & -0.006$ \pm $0.023 \\ 
\hline
4 & 0.0019 & -0.002$ \pm $0.01 \\ 
\hline
5 & 0.0005 & 0.0$ \pm $0.017 \\ 
\hline
6 & 0.0004 & 0.003$ \pm $0.015 \\ 
\hline
7 & -0.0001 & 0.0$ \pm $0.018 \\ 
\hline
8 & -0.0012 & 0.0$ \pm $0.016 \\ 
\hline
9 & -0.0005 & 0.013$ \pm $0.016 \\ 
\hline
10 & 0.0007 & -0.003$ \pm $0.016 \\ 
\hline
11 & 0.0003 & 0.004$ \pm $0.026 \\ 
\hline
12 & -0.0007 & -0.0$ \pm $0.019 \\ 
\hline
13 & -0.0007 & -0.001$ \pm $0.017 \\ 
\hline
14 & -0.0001 & 0.001$ \pm $0.013 \\ 
\hline
15 & 0.0008 & -0.011$ \pm $0.019 \\ 
\hline
16 & -0.0005 & 0.036$ \pm $0.02 \\ 
\hline
17 & -0.0013 & 0.0$ \pm $0.014 \\ 
\hline
18 & 0.0004 & -0.0$ \pm $0.021 \\ 
\hline
19 & 0.0013 & -0.001$ \pm $0.025 \\ 
\hline
20 & -0.0004 & -0.0$ \pm $0.015 \\ 
\hline
21 & -0.0015 & -0.0$ \pm $0.02 \\ 
\hline
22 & -0.0003 & 0.0$ \pm $0.014 \\ 
\hline
23 & 0.0004 & -0.001$ \pm $0.029 \\ 
\hline
24 & -0.0001 & -0.0$ \pm $0.018 \\ 
\hline
25 & 0.0002 & 0.0$ \pm $0.031 \\ 
\hline
26 & -0.0002 & 0.003$ \pm $0.014 \\ 
\hline
27 & 0.0007 & -0.001$ \pm $0.017 \\ 
\hline
28 & 0.0349 & -0.0$ \pm $0.022 \\ 
\hline
29 & 0.9885 & 0.951$ \pm $0.018 \\ 
\hline
30 & 0.112 & 0.336$ \pm $0.017 \\ 
\hline
31 & 0.1765 & 0.359$ \pm $0.019 \\ 
\hline
32 & 0.1004 & 0.269$ \pm $0.019 \\ 
\hline
33 & 0.1384 & 0.319$ \pm $0.019 \\ 
\hline
34 & 0.1248 & 0.254$ \pm $0.017 \\ 
\hline
35 & 0.1312 & 0.345$ \pm $0.017 \\ 
\hline
36 & 0.1302 & 0.288$ \pm $0.024 \\ 
\hline
37 & 0.1289 & 0.271$ \pm $0.018 \\ 
\hline
38 & 0.1292 & 0.324$ \pm $0.013 \\ 
\hline
39 & 0.1296 & 0.315$ \pm $0.015 \\ 
\hline
40 & 0.1298 & 0.297$ \pm $0.016 \\ 
\hline
41 & 0.1256 & 0.306$ \pm $0.017 \\ 
\hline
42 & 0.1264 & 0.332$ \pm $0.014 \\ 
\hline
43 & 0.1309 & 0.338$ \pm $0.019 \\ 
\hline
44 & 0.1268 & 0.312$ \pm $0.016 \\ 
\hline
45 & 0.1246 & 0.348$ \pm $0.017 \\ 
\hline
46 & 0.1232 & 0.349$ \pm $0.015 \\ 
\hline
47 & 0.1257 & 0.327$ \pm $0.017 \\ 
\hline
48 & 0.1241 & 0.312$ \pm $0.017 \\ 
\hline
49 & 0.121 & 0.287$ \pm $0.021 \\ 
\hline
50 & 0.1237 & 0.313$ \pm $0.017 \\ 
\hline
51 & 0.1231 & 0.284$ \pm $0.019 \\ 
\hline
52 & 0.1229 & 0.31$ \pm $0.014 \\ 
\hline
53 & 0.1231 & 0.302$ \pm $0.021 \\ 
\hline
54 & 0.1246 & 0.305$ \pm $0.013 \\ 
\hline
55 & 0.1255 & 0.335$ \pm $0.023 \\ 
\hline
56 & 0.126 & 0.344$ \pm $0.021 \\ 
\hline
57 & 0.127 & 0.309$ \pm $0.015 \\ 
\hline
58 & 0.1267 & 0.343$ \pm $0.013 \\ 
\hline
59 & 0.1289 & 0.287$ \pm $0.022 \\ 
\hline
60 & 0.1322 & 0.319$ \pm $0.016 \\ 
\hline
61 & 0.1329 & 0.334$ \pm $0.023 \\ 
\hline
62 & 0.1279 & 0.265$ \pm $0.015 \\ 
\hline
63 & 0.1303 & 0.312$ \pm $0.02 \\ 
\hline
64 & 0.1296 & 0.332$ \pm $0.011 \\ 
\hline
65 & 0.1236 & 0.333$ \pm $0.016 \\ 
\hline
66 & 0.1386 & 0.31$ \pm $0.026 \\ 
\hline
67 & 0.1009 & 0.272$ \pm $0.018 \\ 
\hline
68 & 0.17 & 0.349$ \pm $0.022 \\ 
\hline
69 & 0.1028 & 0.278$ \pm $0.016 \\ 
\hline
70 & 0.9896 & 0.909$ \pm $0.016 \\ 
\hline
71 & 0.0359 & -0.004$ \pm $0.019 \\ 
\hline
72 & 0.0017 & 0.001$ \pm $0.016 \\ 
\hline
73 & -0.0019 & 0.0$ \pm $0.017 \\ 
\hline
74 & -0.0027 & 0.0$ \pm $0.019 \\ 
\hline
75 & -0.0007 & -0.003$ \pm $0.023 \\ 
\hline
76 & 0.0029 & 0.002$ \pm $0.018 \\ 
\hline
77 & 0.0011 & -0.003$ \pm $0.023 \\ 
\hline
78 & -0.0005 & -0.002$ \pm $0.018 \\ 
\hline
79 & -0.0018 & -0.004$ \pm $0.024 \\ 
\hline
80 & -0.0008 & 0.0$ \pm $0.02 \\ 
\hline
81 & 0.0009 & 0.005$ \pm $0.016 \\ 
\hline
82 & 0.0 & 0.004$ \pm $0.018 \\ 
\hline
83 & 0.0001 & -0.001$ \pm $0.014 \\ 
\hline
84 & -0.0004 & -0.001$ \pm $0.016 \\ 
\hline
85 & -0.0005 & 0.0$ \pm $0.019 \\ 
\hline
86 & 0.0005 & -0.0$ \pm $0.018 \\ 
\hline
87 & 0.0001 & -0.0$ \pm $0.019 \\ 
\hline
88 & 0.0006 & 0.001$ \pm $0.013 \\ 
\hline
89 & 0.0012 & 0.001$ \pm $0.015 \\ 
\hline
90 & -0.001 & 0.006$ \pm $0.015 \\ 
\hline
91 & -0.0006 & 0.001$ \pm $0.015 \\ 
\hline
92 & 0.0018 & 0.001$ \pm $0.018 \\ 
\hline
93 & 0.0 & -0.0$ \pm $0.013 \\ 
\hline
94 & -0.0029 & 0.007$ \pm $0.016 \\ 
\hline
95 & -0.0014 & 0.013$ \pm $0.02 \\ 
\hline
96 & -0.0015 & 0.002$ \pm $0.02 \\ 
\hline
97 & -0.0026 & -0.0$ \pm $0.018 \\ 
\hline
98 & -0.0016 & 0.004$ \pm $0.012 \\ 
\hline
99 & 0.0001 & -0.0$ \pm $0.018 \\ 
\hline
100 & -0.0005 & -0.0$ \pm $0.014 \\ 
\hline
101 & -0.0013 & -0.002$ \pm $0.016 \\ 
\hline
102 & 0.0005 & -0.0$ \pm $0.011 \\ 
\hline
103 & -0.0001 & -0.011$ \pm $0.014 \\ 
\hline
\caption{The expectation of $\hat{P}_{0,x} \hat{P}_{1,x+1} + \hat{P}_{1,x} \hat{P}_{0,x+1}$ for $t=5.0$ displayed in Fig.\ref{fig:quarksIBM104}.}
\end{longtable}

\end{document}